\documentclass[preprint,journal]{vgtc}       

\ifpdf
  \pdfoutput=1\relax                   
  \usepackage{graphicx}                
  \DeclareGraphicsExtensions{.pdf,.png,.jpg,.jpeg} 
\else
  \usepackage{graphicx}                
  \DeclareGraphicsExtensions{.eps}     
\fi%

\graphicspath{{figures/}{pictures/}{images/}{./}} 

\usepackage{microtype}                 
\PassOptionsToPackage{warn}{textcomp}  
\usepackage{textcomp}                  
\usepackage{mathptmx}                  
\usepackage{times}                     
\usepackage{cite}                      
\usepackage{tabu}                      
\usepackage{booktabs}                  

\usepackage{amsmath}
\usepackage{amssymb}
\DeclareMathAlphabet{\mathcal}{OMS}{cmsy}{m}{n}

\usepackage{algorithm}
\usepackage{algpseudocode}

\usepackage[normalem]{ulem}

\usepackage{color}
\usepackage[dvipsnames]{xcolor}

\usepackage{enumitem}

\usepackage{multirow}

\newcommand{\removed}[1]{\textcolor{red}{\sout{#1}}}

\ifx\cleanversion\undefined
\else

  \renewcommand{\removed}[1]{\iffalse#1\fi}
\fi

\ieeedoi{10.1109/TVCG.2019.2934312}

\onlineid{1048}

\vgtccategory{Research}
\vgtcpapertype{algorithm/technique}

\title{InSituNet: Deep Image Synthesis for Parameter Space Exploration of Ensemble Simulations}

\author{Wenbin He, Junpeng Wang, Hanqi Guo, \textit{Member, IEEE}, Ko-Chih Wang, Han-Wei Shen, \textit{Member, IEEE},\\ Mukund Raj, Youssef S. G. Nashed, and Tom Peterka, \textit{Member, IEEE}}
\authorfooter{
\item Wenbin He, Junpeng Wang, Ko-Chih Wang, and Han-Wei Shen are with the Department of Computer Science and Engineering, The Ohio State University. E-mail: \{he.495, wang.7665, wang.3182, shen.94\}@osu.edu.
\item Hanqi Guo, Mukund Raj, Youssef S. G. Nashed, and Tom Peterka are with the Mathematics and Computer Science Division, Argonne National Laboratory. E-mail: \{hguo, mraj, ynashed, tpeterka\}@anl.gov.
}

\abstract{
We propose InSituNet, a deep learning based surrogate model to support parameter space exploration for ensemble simulations that are visualized in situ.  In situ visualization, generating visualizations at simulation time, is becoming prevalent in handling large-scale simulations because of the I/O and storage constraints.  However, in situ visualization approaches limit the flexibility of post-hoc exploration because the raw simulation data are no longer available.  Although multiple image-based approaches have been proposed to mitigate this limitation, those approaches lack the ability to explore the simulation parameters.  Our approach allows flexible exploration of parameter space for large-scale ensemble simulations by taking advantage of the recent advances in deep learning.  Specifically, we design InSituNet as a convolutional regression model to learn the mapping from the simulation and visualization parameters to the visualization results.  With the trained model, users can generate new images for different simulation parameters under various visualization settings, which enables in-depth analysis of the underlying ensemble simulations.  We demonstrate the effectiveness of InSituNet in combustion, cosmology, and ocean simulations through quantitative and qualitative evaluations.
} 


\CCScatlist{ 
 \CCScat{K.6.1}{Management of Computing and Information Systems}%
{Project and People Management}{Life Cycle};
 \CCScat{K.7.m}{The Computing Profession}{Miscellaneous}{Ethics}
}

\vgtcinsertpkg

\begin{document}



\maketitle

\section{Introduction}
\label{sec:introduction}

\setlength{\abovedisplayskip}{3pt}
\setlength{\belowdisplayskip}{3pt}

Ensemble simulations~\cite{WangHLS18} have been playing an increasingly important role in various scientific and engineering disciplines, such as computational fluid dynamics, cosmology, and weather research.  As the computational power of modern supercomputers continues to grow, ensemble simulations are more often conducted with a large number of parameter settings in high spatial and/or temporal resolutions.
Despite the advances in accuracy and reliability of simulation results, however, two challenges have emerged: (1) I/O bottleneck for the movement of the large-scale simulation data and (2) effective exploration and analysis of the simulation parameters.
In situ visualization~\cite{Ma09, BauerAACGKMOVWB16}, which generates visualization at simulation time and stores only the visualization results (that are much smaller than the raw simulation data~\cite{AhrensJOPRP14, AhrensJOPRFBPSB14}) for post-hoc analysis, addresses the first challenge to some extent.  However, it also limits the flexibility of post-hoc exploration and analysis, because the raw simulation data are no long available.

This study focuses on improving scientists' ability in exploring the in situ visualization results of ensemble simulations and extending their capability in investigating the influence of different simulation parameters.
Several pioneering works have been proposed to facilitate post-hoc exploration of in situ visualization results.  For example, the Cinema framework~\cite{AhrensJOPRP14, AhrensJOPRFBPSB14} visualized the simulation data from different viewpoints in situ and collected images to support post-hoc exploration.  The volumetric depth images~\cite{FreySE13, FernandesFSE14} stored ray segments with composited color and opacity values to enable post-hoc exploration of arbitrary viewpoints for volume rendering.
However, these approaches focus more on extending the capability to explore the visual mapping parameters (e.g., transfer functions) and view parameters (e.g., view angles) and have little consideration of the simulation parameters, which are important in studying ensemble simulations.

Simulation parameter space exploration is not trivial, because the relationship between the simulation parameters and outputs is often highly complex.  The majority of existing simulation parameter space exploration approaches~\cite{WangLSL17, BrucknerM10} resorted to visualizing a set of simulation parameters and outputs simultaneously and revealing the correspondence between the parameters and outputs through visual linkings.  However, these approaches often depend on the raw simulation data that might not be available for large-scale ensemble simulations.  Moreover, these approaches have limited ability in inferring simulation outputs with respect to new parameters.  Hence, extra simulations have to be conducted for new parameters, which cost enormous computational resources for most scientific simulations.

In this work, we propose InSituNet, a deep learning based surrogate model to support parameter space exploration for ensemble simulations that are visualized in situ.
Our work is based on the observation that images of high accuracy and fidelity can be generated with deep neural networks for various image synthesis applications, such as super-resolution~\cite{DongLHT16, JohnsonAL16, LedigTHCCAATTWS17}, inpainting~\cite{XieXC12, PathakKDDE16}, texture synthesis~\cite{GatysEB15a, ZhouZBLOH18}, and rendering~\cite{DosovitskiySB15, BergerLL19}.  Specifically, we train InSituNet to learn the end-to-end mapping from the simulation, visual mapping, and view parameters to visualization images.
The trained model enables scientists to interactively explore synthesized visualization images for different simulation parameters under various visualization settings without actually executing the expensive simulations.
Our approach consists of three major steps.
\begin{enumerate}
[topsep=0.2em]
\itemsep-0.2em
\item \textbf{In situ training data collection from ensemble simulations}\quad
Given ensemble simulations conducted with different simulation parameters, we visualize the generated simulation data in situ with various visual mapping and view parameters.  The resulting visualization images and the corresponding parameters are collected and used for the offline training of InSituNet.
\item \textbf{Offline training of InSituNet}\quad Given the parameters and image pairs, we train InSituNet (i.e., a convolutional regression model) with cutting-edge deep learning techniques on image synthesis to map simulation, visual mapping, and view parameters to visualization images directly.
\item \textbf{Interactive post-hoc exploration and analysis}\quad With the trained InSituNet, we build an interactive visual interface that enables scientists to explore and analyze the simulation from two perspectives: (1) inferring visualization results for arbitrary parameter settings within the parameter space with InSituNet's forward propagations and (2) analyzing the sensitivity of different parameters with InSituNet's backward propagations.
\end{enumerate}

We demonstrate the effectiveness of the proposed approach in combustion, cosmology, and ocean simulations, and compare the predicted images of InSituNet with the ground truth and alternative methods.  In addition, we evaluate the influence of different hyperparameters of InSituNet (e.g., the choice of loss functions and the network architectures) and provide guidance in configuring the hyperparameters.  In summary, the contributions of this paper are threefold:
\begin{itemize}
[topsep=0.2em]
\itemsep-0.2em
  \item A deep image synthesis model (i.e., InSituNet) that enables post-hoc parameter space exploration of ensemble simulations
  \item An interactive visual interface to explore and analyze the parameters of ensemble simulations with the trained InSituNet
  \item A comprehensive study revealing the effects of different hyperparameters of InSituNet and providing guidance for applying InSituNet to other simulations
\end{itemize}

\section{Related Work}
\label{sec:related_work}

In this section, we review related work in image-based in situ visualization, parameter space exploration of ensemble simulations, and deep learning for visualization.

\subsection{Image-Based In Situ Visualization}

Based on the output, in situ visualization can be categorized into image-based~\cite{AhrensJOPRP14, AhrensJOPRFBPSB14}, distribution-based~\cite{DuttaCHSC17}, compression-based~\cite{doi:10.1002/cpe.2887, conf/ipps/DiC16}, and feature-based~\cite{BremerWTPDB11} approaches.  We regard our work as an image-based approach, which visualizes simulation data in situ and stores images for post-hoc analysis.  Tikhonova et al.~\cite{TikhonovaCM10a, TikhonovaCM10b, TikhonovaCM10c} generated images of multiple layers in situ to enable the adjustment of transfer functions in post-hoc analysis.  Frey et al.~\cite{FreySE13} proposed volumetric depth images, a compact representation of volumetric data that can be rendered efficiently with arbitrary viewpoints.  Fernandes et al.~\cite{FernandesFSE14} later extended volumetric depth images to handle time-varying volumetric data.  Biedert and Garth~\cite{BiedertG15} combined topology analysis and image-based data representation to preserve flexibility for post-hoc exploration and analysis.  Ahrens et al.~\cite{AhrensJOPRP14, AhrensJOPRFBPSB14} proposed Cinema, a framework that stores visualization images in situ and performs post-hoc analysis via exploration and composition of those images.

Compared with these approaches, our work supports not only the exploration of various visual mapping and view parameters but also the creation of visualizations under new simulation parameters without actually running the simulation.

\subsection{Parameter Space Exploration}
The existing parameter space exploration works for ensemble simulations can generally be reviewed from two perspectives, the adopted visualization techniques and the objective of parameter space exploration.
Visualization techniques that designed for high-dimensional data are often borrowed to visualize the parameter space of ensemble simulations, as the simulation parameters are typically treated as multidimensional vectors. These techniques include but are not limited to: parallel coordinate plots~\cite{ObermaierBJ16, WangLSL17}, radial plots~\cite{CoffeyLEK13, HaidongZCMZMLQ15, BrucknerM10}, scatter plots~\cite{MatkovicGKH09, SplechtnaMGJH15, OrbanKBAR19}, line charts~\cite{BiswasLLS17}, matrices~\cite{PocoDWHSHCBS14}, and glyphs~\cite{BockPMRRY15}.
For the objectives of parameter space exploration, we believe the six tasks sorted out by Sedlmair et al.~\cite{sedlmair2014visual} could best summarize the literature, which are optimization~\cite{torsney2011tuner}, partitioning~\cite{WangLSL17, bergner2013paraglide}, filtering~\cite{piringer2010hypermoval}, outliers~\cite{piringer2010hypermoval, potter2009ensemble}, uncertainty~\cite{booshehrian2012vismon, BiswasLLS17}, and sensitivity~\cite{BiswasLLS17}. We refer the interested readers to the work of Sedlmair et al.~\cite{sedlmair2014visual} for the detailed definition of each task, as well as the example visualization works.

The aforementioned parameter visualization techniques and analysis tasks mostly focus on a limited number of simulation inputs and outputs collected from ensemble runs. In this paper, we train a surrogate model to extend our study to arbitrary parameter settings within the parameter space, even if the simulations were not executed with those settings. In addition, our approach is incorporated with in situ visualization, which is widely used in large-scale ensemble simulations.

\subsection{Deep Learning for Visualization}

The visualization community has started to incorporate deep learning in visualization research.  For example, Hong et al.~\cite{HongZY18} used long short-term memory~\cite{HochreiterS97} to estimate access pattern for parallel particle tracing.  Han et al.~\cite{HanTW18} used autoencoders~\cite{RumelhartHW86} to cluster streamlines and streamsurfaces.  Xie et al.~\cite{XieXM19} used neural network embeddings to detect anomalous executions in high performance computing applications.
Berger et al.~\cite{BergerLL19} proposed a deep learning approach to assist transfer function design using generative adversarial networks (GANs)~\cite{GoodfellowAMXFOCB14}, which is closely related to our approach.  Specifically, we focus on parameter space exploration of ensemble simulations instead of transfer function design for volume rendering.

Our work is related to deep learning based image synthesis, which has been used in various applications, including super-resolution~\cite{DongLHT16, JohnsonAL16, LedigTHCCAATTWS17}, denoising~\cite{XieXC12, ZhangZCMZ17}, inpainting~\cite{XieXC12, PathakKDDE16}, texture synthesis~\cite{GatysEB15a, ZhouZBLOH18}, text-to-image synthesis~\cite{ReedAYLSL16}, style transfer~\cite{GatysEB15b, JohnsonAL16, ZhuPIE17}, and rendering~\cite{DosovitskiySB15, BergerLL19}.  We investigate and combine different state-of-the-art deep learning techniques on image synthesis (e.g., perpetual losses~\cite{JohnsonAL16, LedigTHCCAATTWS17} and GANs~\cite{GoodfellowAMXFOCB14}) to improve the quality of our image synthesis results.

\section{Overview}
\label{sec:overview}
\vspace{-10pt}

\setlength{\abovecaptionskip}{4pt}
\setlength{\belowcaptionskip}{-10pt}
\begin{figure}[tbh]
  \centering
  \includegraphics[width=\linewidth]{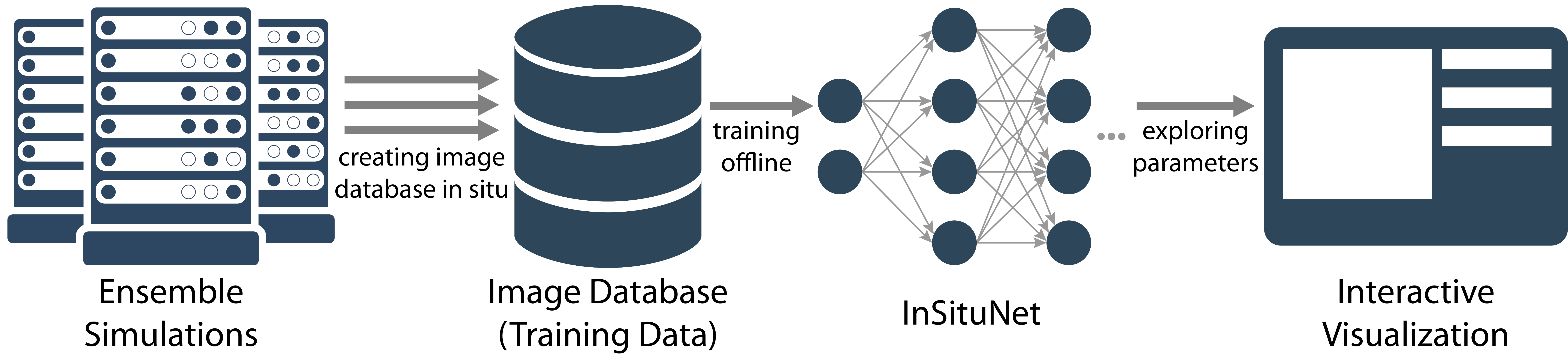}
  \caption{Workflow of our approach.  Ensemble simulations are conducted with different simulation parameters on supercomputers, and visualization images are generated in situ for different visual mapping and view parameters.  The generated images and the parameters are collected into an image database.  A deep image synthesis model (i.e., InSituNet) is then trained offline based on the collected data, which is later used for parameter space exploration through an interactive visual interface.
  }
  \label{fig:overview}
\end{figure}
\setlength{\abovecaptionskip}{10pt}
\setlength{\belowcaptionskip}{0pt}

Figure~\ref{fig:overview} provides the workflow of our approach, which consists of three major components.  First, given ensemble simulations conducted with different simulation parameters, we visualize the generated simulation outputs in situ with different visual mapping and view parameters on supercomputers.  The three groups of parameters---simulation, visual mapping, and view parameters---along with the corresponding visualization results (i.e., images) are collected to constitute an image database (Section~\ref{sec:data_coll}).  Second, with the collected data pairs between parameters and the corresponding images, we train InSituNet to learn the end-to-end mapping from the simulation inputs to the visualization outputs (Section~\ref{sec:dl_model}).  To improve the accuracy and fidelity of the generated images, we use and combine different state-of-the-art deep learning techniques on image synthesis.  Third, with the trained InSituNet, we build an interactive visual interface (Section~\ref{sec:vis_sys}) to explore and analyze the parameters from two aspects: (1) predicting visualization images interactively for arbitrary simulation, visual mapping, and view parameters within the parameter space and (2) investigating the sensitivity of different input parameters to the visualization results.

\section{In Situ Training Data Collection}
\label{sec:data_coll}
\vspace{-10pt}

\setlength{\abovecaptionskip}{4pt}
\setlength{\belowcaptionskip}{-10pt}
\begin{figure}[tbh]
  \centering
  \includegraphics[width=.8\linewidth]{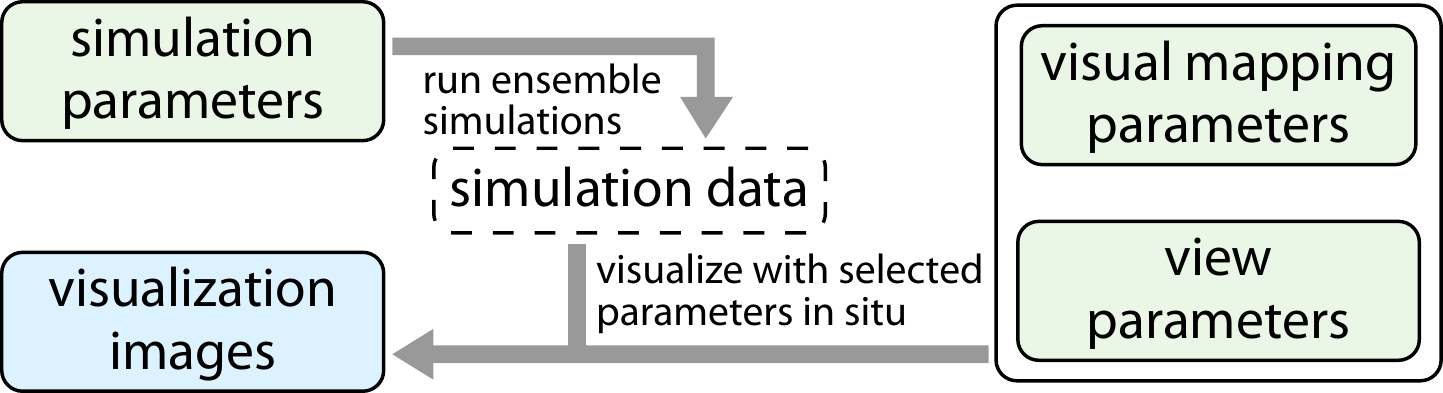}
  \caption{Our in situ training data collection pipeline.
  Simulation data, generated with different simulation parameters, are visualized in situ with different visual mapping and view parameters.
  The in situ visualization generates a large number of images, which are collected along with the corresponding parameters for the training of InSituNet offline.
  }
  \label{fig:data_coll}
\end{figure}
\setlength{\abovecaptionskip}{10pt}
\setlength{\belowcaptionskip}{0pt}

Figure~\ref{fig:data_coll} illustrates our in situ training data collection pipeline.
Given ensemble simulations conducted with different simulation parameters, we perform in situ visualization with a desired set of visual mapping parameters (e.g., isosurfaces extraction with a set of isovalues) and different view parameters (e.g., viewpoints).  We denote an instance of simulation, visual mapping, and view parameters as $P_{sim}$, $P_{vis}$, and $P_{view}$, respectively, which corresponds to a visualization image $I$.
The parameters (highlighted in green in Figure~\ref{fig:data_coll}) and the corresponding visualization images (highlighted in blue in Figure~\ref{fig:data_coll}) constitute data pairs, which will be stored and used to train InSituNet.
InSituNet learns a function $\mathcal{F}$ that maps the three groups of parameters to the corresponding visualization image, which can be defined as
\begin{equation}
\label{eq:mapping}
\mathcal{F}(P_{sim},P_{vis},P_{view})\rightarrow{I},
\end{equation}
so that it can predict visualization images for unseen parameters.
In the following, we discuss the three groups of parameters in detail.

\textbf{Simulation parameters $P_{sim}$} are represented as a vector with one or more dimensions, and the value range of each dimension is defined by scientists.  By sweeping the parameters within the defined ranges, ensemble simulations are conducted to generate the ensemble data.

\textbf{Visual mapping parameters $P_{vis}$} are predefined operations to visualize the generated simulation data, such as pseudo-coloring with predefined color schemes.  Note that we limit the users' ability in selecting arbitrary visual mappings to produce and store fewer images.

\textbf{View parameters $P_{view}$} are used to control the viewpoints that the images are created from.  In this work, we define the viewpoints by a camera rotating around the simulation data, which is controlled by azimuth $\theta\in[0, 360]$ and elevation ${\phi}\in[-90, 90]$.  For panning and zooming, we resort to image-based operations (i.e., panning and resizing the images) as proposed in~\cite{AhrensJOPRP14}). To train a deep learning model that can predict visualization images for arbitrary viewpoints, we sample the azimuth and elevation and generate images from the sampled viewpoints.  Based on our study, we found that taking 100 viewpoints for each ensemble member is sufficient to train InSituNet.

With the specified values for the three groups of parameters, we generate the corresponding visualization images.  Our work uses RGB images compressed to the portable network graphics (PNG) format instead of more sophisticated image formats, such as volumetric depth images~\cite{FreySE13, FernandesFSE14} or explorable images~\cite{TikhonovaCM10a, TikhonovaCM10b, TikhonovaCM10c}, for two reasons.
First, the benefits of using those sophisticated image formats, such as supporting changing of viewpoints, can be achieved by InSituNet trained on the RGB images.
Second, RGB images are more generally applicable for various visualizations and more easily to be handled by neural networks compared with those sophisticated image formats.

\section{InSituNet Architecture and Training}
\label{sec:dl_model}
\vspace{-10pt}

\setlength{\abovecaptionskip}{4pt}
\setlength{\belowcaptionskip}{-10pt}
\begin{figure}[tbh]
  \centering
  \includegraphics[width=\linewidth]{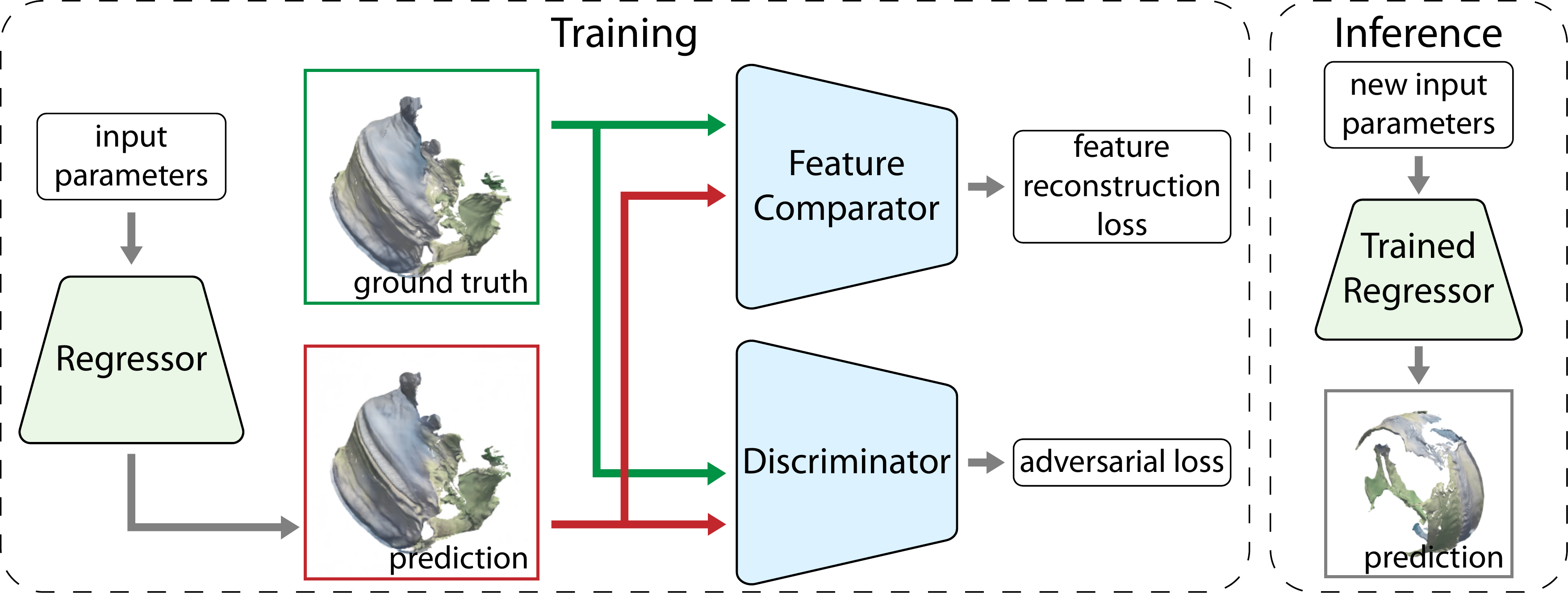}
  \caption{Overview of InSituNet, which is a convolutional regression model that predicts visualization images from input parameters.  During training, the regression model is trained based on the losses computed with the assist of a pretrained feature comparator and a discriminator.
  }
  \label{fig:nn_overview}
\end{figure}
\setlength{\abovecaptionskip}{10pt}
\setlength{\belowcaptionskip}{0pt}

Figure~\ref{fig:nn_overview} illustrates the training and inference pipelines of InSituNet.  In the training stage, InSituNet consists of three subnetworks: a regressor, a feature comparator, and a discriminator.  The regressor $R_{\omega}$ is a deep neural network (defined by a set of weights $\omega$) modeling the function that maps input parameters to visualization images as defined in Equation~\ref{eq:mapping}.  To train a regressor that can generate images of high fidelity and accuracy, we introduced the feature comparator $F$ and the discriminator $D_{\upsilon}$ to compute losses by comparing the predicted and the ground truth images.  The feature comparator is a pretrained neural network whose convolutional kernels are used to extract and compare image features (e.g., edges, shapes) between the predicted and the ground truth images to obtain a feature reconstruction loss.  The discriminator $D_{\upsilon}$ is a deep neural network whose weights $\upsilon$ are updated during training to estimate the divergence between the distributions of the predicted and the ground truth images.  The divergence is known as the adversarial loss~\cite{GoodfellowAMXFOCB14}, which is combined with the feature reconstruction loss to train $R_{\omega}$.  In the inference stage, we need only the trained $R_{\omega}$, which can predict visualization images for parameters that are not in the training data.  In the following, we discuss the network architecture, the loss function, and the training process in detail.


\subsection{Network Architecture}
\label{sec:architecture}

Three subnetworks are involved during training: the regressor $R_{\omega}$, feature comparator $F$, and discriminator $D_{\upsilon}$.  The regressor $R_{\omega}$ and discriminator $D_{\upsilon}$ are two deep residual convolutional neural networks~\cite{HeZRS16} parameterized by the weights $\omega$ and $\upsilon$, respectively.  The architectures of $R_{\omega}$ and $D_{\upsilon}$ are designed by following the network architecture proposed by~\cite{MiyatoKKY18, KurachLZMG18}, because the scale of our image synthesis problem is similar to theirs.  For the feature comparator $F$, we use the pretrained VGG-19 model~\cite{SimonyanZ15}, which has been widely used in many deep image synthesis approaches~\cite{JohnsonAL16, LedigTHCCAATTWS17, HouSSQ17}.

\subsubsection{Regressor $R_{\omega}$}
\vspace{-10pt}

\setlength{\abovecaptionskip}{4pt}
\setlength{\belowcaptionskip}{-10pt}
\begin{figure}[tbh]
  \centering
  \includegraphics[width=\linewidth]{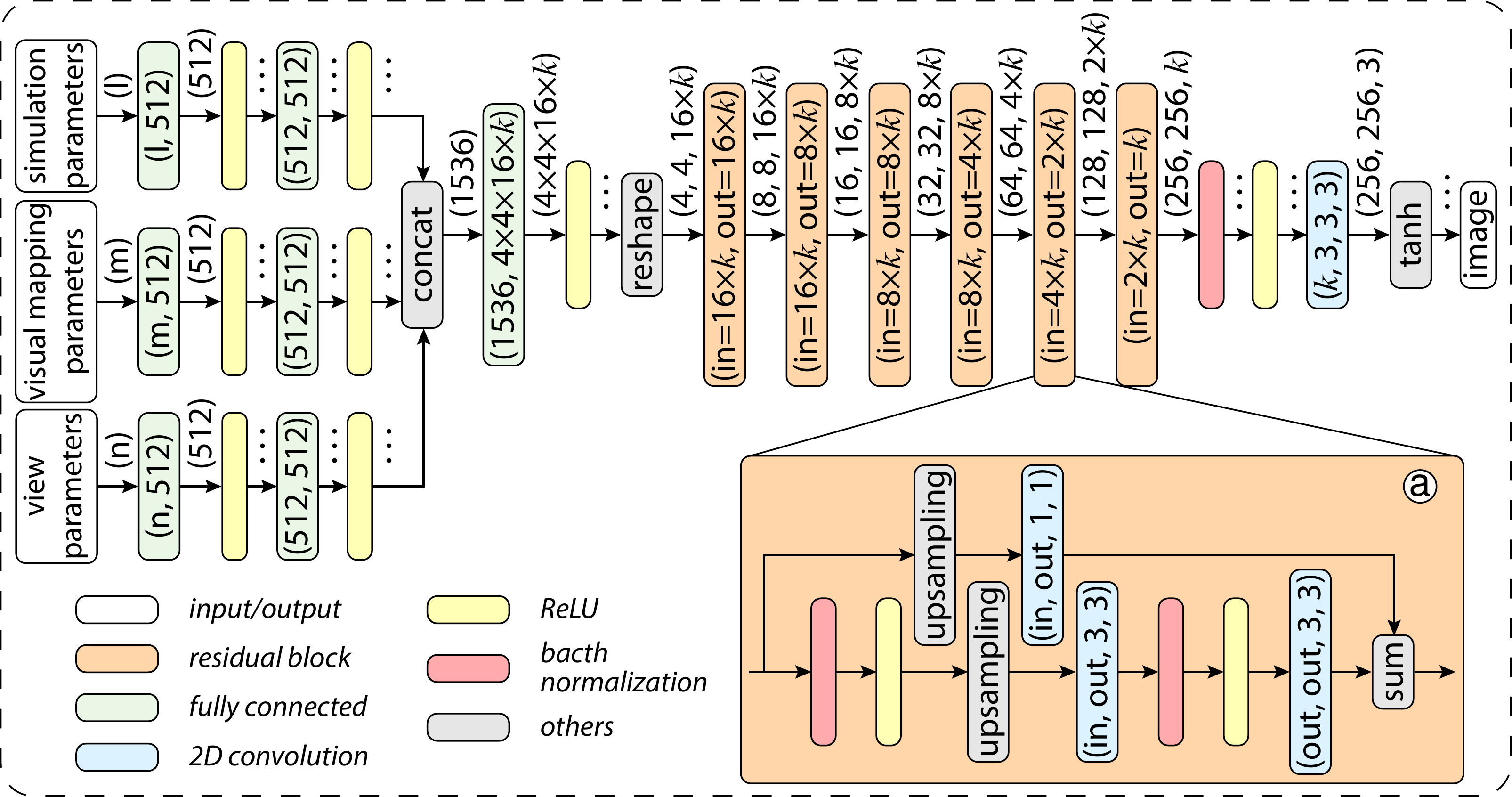}
  \caption{Architecture of $R_{\omega}$, which encodes input parameters into a latent vector with fully connected layers and maps the latent vector into an output image with residual blocks.  The size of $R_{\omega}$ is defined by $k$, which controls the number of convolutional kernels in the intermediate layers.}
  \label{fig:generator}
\end{figure}
\setlength{\abovecaptionskip}{10pt}
\setlength{\belowcaptionskip}{0pt}

The architecture of $R_{\omega}$ is shown in Figure~\ref{fig:generator}, which takes the $P_{sim}$, $P_{vis}$, and $P_{view}$ as inputs and outputs a predicted image $I$.  The three types of parameters are first fed into three groups of fully connected layers separately, and the outputs are then concatenated and fed into another fully connected layer to encode them into a latent vector.  Note that the parameters could also be concatenated first and then fed into fully connected layers.  However, as each parameter is fully connected with all neurons in the next layer, more weights will be introduced in the network and the network size will increase.  Next, the latent vector is reshaped into a low-resolution image, which is mapped to a high-resolution output image through residual blocks performing 2D convolutions and upsamplings.  Following the commonly used architecture~\cite{MiyatoKKY18, KurachLZMG18}, we use the rectified linear unit (\texttt{ReLU}) activation function~\cite{NairH10} in all layers except the output layer.  For the output layer, we use the \texttt{tanh} function to normalize each pixel into $\lbrack-1, 1\rbrack$.

Note that we introduce a constant $k$ in the network architecture to control the number of convolutional kernels in the intermediate layers.  The constant $k$ is used to balance the expressive power and the size and training time of $R_\omega$ to cope with datasets in different complexities.

\textbf{Residual Blocks}\quad  $R_\omega$ consists of several residual blocks (Figure~\ref{fig:generator}a), which are proposed in~\cite{HeZRS16} to improve the performance of neural networks with increasing depth.  We adopted the residual blocks here because $R_\omega$ often needs to be very deep (i.e., more than 10 convolutional layers) to synthesize images with high-resolutions.
Inside each residual block, the input image is first upsampled by using nearest neighbor upsampling.  The upsampled image is then fed into two convolutional layers with kernel size $3{\times}3$.  In the end, the original input image is added to the output, and the result is sent to the next layer.  Batch normalizations are performed on the output of each convolutional layer to stabilize the training.  Note that if the resolution or the channel number of the input image is not the same as the output, we perform the upsampling and convolution operations on the input image to transform it into the size of the output.

\subsubsection{Discriminator $D_{\upsilon}$}

\setlength{\abovecaptionskip}{4pt}
\setlength{\belowcaptionskip}{-18pt}
\begin{figure}[tb]
  \centering
  \includegraphics[width=\linewidth]{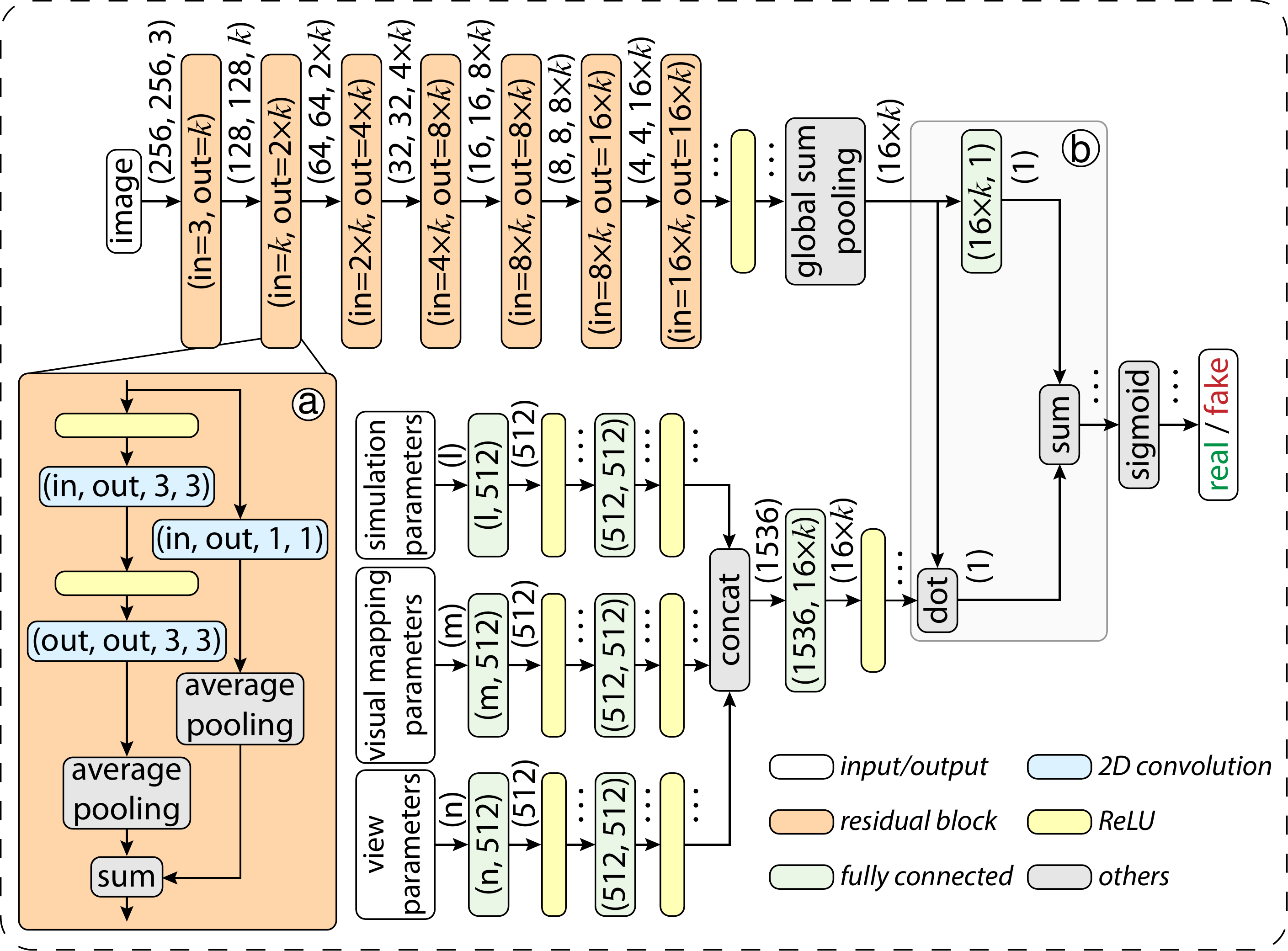}
  \caption{Architecture of $D_{\upsilon}$.  Input parameters and the predicted/ground truth image are transformed into latent vectors with fully connected layers and residual blocks, respectively.  The latent vectors are then incorporated by using the projection-based method~\cite{MiyatoK18} to predict how likely the image is a ground truth image conditioning on the given parameters.  Similar to $R_{\omega}$, the size of $D_{\upsilon}$ is controlled by the constant $k$.}
  \label{fig:discriminator}
\end{figure}
\setlength{\abovecaptionskip}{10pt}
\setlength{\belowcaptionskip}{0pt}

The architecture of $D_{\upsilon}$ is shown in Figure~\ref{fig:discriminator}, which takes a predicted/ground truth image and the corresponding parameters as inputs and produces a likelihood value indicating how likely the input image is a ground truth image conditioning on the given parameters.  With the likelihood values, an adversarial loss can be defined to update $D_{\upsilon}$ and $R_{\omega}$ (details in Section~\ref{sec:advloss}).
Similar to $R_{\omega}$, the three types of parameters are encoded into a latent vector in $D_{\upsilon}$ through fully connected layers.  Meanwhile, the input image is fed through several residual blocks to derive its intermediate representation that is a latent vector.  The latent vectors are then incorporated to obtain the likelihood value conditioning on the three groups of parameters.  \texttt{ReLU} activations are used in all layers except the output layer, which instead uses the \texttt{sigmoid} function to derive a likelihood value within $\lbrack0, 1\rbrack$.

\textbf{Residual blocks}\quad  The architecture of the residual blocks in $D_{\upsilon}$ (Figure~\ref{fig:discriminator}a) is similar to that in $R_{\omega}$ except that downsampling (average pooling in this work) is performed instead of upsampling to transform images into low-resolution representations and no batch normalization is performed, because it often hurts the performance of $D_{\upsilon}$~\cite{LucicKMGB18, KurachLZMG18}.

\textbf{Projection-based condition incorporation}\quad  We employed the projection-based method~\cite{MiyatoK18} to incorporate the conditional information (i.e., the three groups of parameters) with the input image.  This method computes a dot product between the data to be incorporated, which in our work is the latent vector of the input parameters and the latent vector of the image (Figure~\ref{fig:discriminator}b).
Compared with other condition incorporation methods, such as vector concatenation~\cite{MirzaO14, BergerLL19}, the projection-based method improves the quality of conditional image synthesis results, as demonstrated by Miyato and Koyama~\cite{MiyatoK18}.

\subsubsection{Feature Comparator $F$}
\vspace{-10pt}

\setlength{\abovecaptionskip}{2pt}
\setlength{\belowcaptionskip}{-10pt}
\begin{figure}[tbh]
  \centering
  \includegraphics[width=\linewidth]{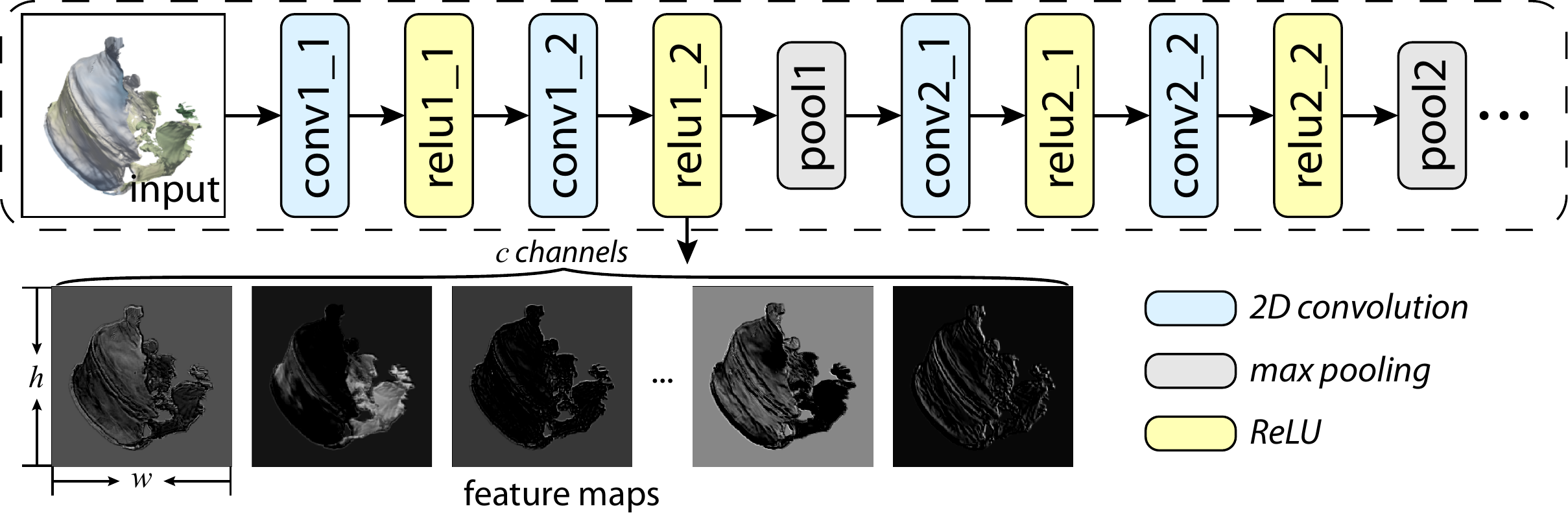}
  \caption{Architecture of $F$ (i.e., VGG-19 network), where each layer is labeled with its name.  Feature maps are extracted through convolutional layers (e.g., \texttt{relu1\_2}) for feature-level comparisons.}
  \label{fig:vgg19}
\end{figure}
\setlength{\abovecaptionskip}{10pt}
\setlength{\belowcaptionskip}{0pt}

To produce high quality image synthesis results, we also strive to minimize the feature-level difference between the generated and the ground truth image by employing a commonly used feature comparator $F$, namely the pretrained VGG-19 model~\cite{SimonyanZ15} shown in Figure~\ref{fig:vgg19}.  $F$ is a convolutional neural network, and the convolutional kernels on each layer have been pretrained to extract certain types of image features, such as edges and shapes.  With it, we extract the features from a generated image, as well as its corresponding ground truth image, and minimize the difference between those features to improve the quality of the generated image (see details in Section~\ref{sec:reconloss}).
Specifically, we use the layer \texttt{relu1\_2} to extract feature maps for comparison based on two observations.  First, early layers such as the layer \texttt{relu1\_2} of the VGG-19 network focus on low-level features such as edges and basic shapes, which commonly exist in scientific visualization images.  Second, through our experiments we found that artifacts are introduced into the generated image by pooling layers (e.g., \texttt{pool1} in Figure~\ref{fig:vgg19}).  Hence, we use the layer \texttt{relu1\_2} that is before the first pooling layer.

\subsection{Loss Function}
\label{sec:loss}

Given an image $\hat{I}$ generated using $R_{\omega}$ and the corresponding ground truth image $I$, a loss function $\mathcal{L}$ is defined by measuring the difference between them.  Minimizing $\mathcal{L}$ can, therefore, be conducted by updating the parameters $\omega$ of $R_{\omega}$ over an iterative training process.  The most straightforward choice for $\mathcal{L}$ is the average of the pixel wise distance between $\hat{I}$ and $I$, such as the mean squared error.  As shown in earlier works~\cite{JohnsonAL16, LedigTHCCAATTWS17, HouSSQ17}, however, the average pixel wise distance often produces over-smoothed images, lacking high-frequency features.

In this work, we define $\mathcal{L}$ by combining two advanced loss functions: a feature reconstruction loss~\cite{JohnsonAL16} $\mathcal{L}_{feat}^{F,l}$ and an adversarial loss~\cite{GoodfellowAMXFOCB14} $\mathcal{L}_{adv\!\_R}$, namely
\begin{equation}
\label{eq:loss_g}
\mathcal{L}=\mathcal{L}_{feat}^{F,l} + \lambda\mathcal{L}_{adv\!\_R},
\end{equation}
where $\lambda$ is the coefficient between them.  $\mathcal{L}_{feat}^{F,l}$ measures the difference between features extracted from the feature comparator $F$ using its convolutional layer $l$, whereas $\mathcal{L}_{adv\!\_R}$ quantifies how easily the discriminator $D_{\upsilon}$ can differentiate the generated images from real ones.  As can be seen, minimizing $\mathcal{L}_{adv\!\_R}$ requires training $R_{\omega}$ and $D_{\upsilon}$ together in an adversarial manner (i.e., the adversarial theory of GANs~\cite{GoodfellowAMXFOCB14}).  In order to train $D_{\upsilon}$, an adversarial loss $\mathcal{L}_{adv\!\_D}$ is used.

\subsubsection{Feature Reconstruction Loss}
\label{sec:reconloss}

The feature reconstruction loss between image $\hat{I}$ and $I$ is defined by measuring the difference between their extracted features~\cite{JohnsonAL16, LedigTHCCAATTWS17, HouSSQ17}.  Specifically, for a given image $I$, our feature comparator $F$ (the pretrained VGG-19) is applied on $I$ and extracts a set of feature maps, denoted as $F^{l}(I)$.  Here $l$ indicates which layer the feature maps are from (e.g., the \texttt{relu1\_2} of $F$).  The extracted feature maps can be considered as a 3D matrix of dimension $h{\times}{w}{\times}{c}$, where $h$, $w$, and $c$ are the height, width, and number of channels, respectively, as shown in Figure~\ref{fig:vgg19}.  The feature reconstruction loss between $\hat{I}$ and $I$ can, therefore, be defined as the pixel wise mean squared error between $F^{l}(\hat{I})$ and $F^{l}(I)$.  Extending this definition to a batch of images, the feature reconstruction loss between $\hat{I}_{0:b-1}$ and $I_{0:b-1}$ ($b$ is the batch size) is
\begin{equation}
\label{eq:loss_perc}
\mathcal{L}_{feat}^{F,l}=\frac{1}{hwcb}\sum_{i=0}^{b-1}\lVert{F^{l}(I_i)-F^{l}(\hat{I}_i)}\rVert_2^2.
\end{equation}

Using the feature reconstruction loss enables our regressor to produce images sharing similar feature maps with the corresponding ground truth images, which lead to images with sharper features.

\subsubsection{Adversarial Loss}
\label{sec:advloss}

In addition to the feature reconstruction loss described above, we add an adversarial loss $\mathcal{L}_{adv\!\_R}$ into the loss function.  Unlike the feature reconstruction loss, which measures the difference between each pair of images, the adversarial loss focuses on identifying and minimizing the divergence between two image distributions following the adversarial theory of GANs.
Specifically, our discriminator $D_{\upsilon}$ is trained along with the regressor $R_{\omega}$ to differentiate images generated by $R_{\omega}$ with ground truth images.  As the regressor $R_{\omega}$ becomes stronger over the training, the discriminator $D_{\upsilon}$ is forced to identify more subtle differences between the generated images and the ground truth.

The adversarial loss can be used as complementary to the feature reconstruction loss for two reasons.  First, the feature reconstruction loss focuses on the average difference between images, and the adversarial loss focuses on local features that are the most important to differentiate the predicted and ground truth images.  Second, the feature reconstruction loss compares the difference between each pair of the generated and ground truth images, and the adversarial loss measures divergence between two image distributions.

In this work, we use the standard adversarial loss presented in~\cite{GoodfellowAMXFOCB14}, which uses different loss functions for the generator and discriminator.  For the generator (i.e., our regressor $R_{\omega}$), the adversarial loss is
\begin{equation}
\label{eq:loss_adv_g}
\mathcal{L}_{adv\!\_R}=-\frac{1}{b}\sum_{i=0}^{b-1}\log{D_{\upsilon}(\hat{I}_i)},
\end{equation}
which reaches the minimum when the discriminator cannot differentiate the generated images from the ground truth images.  This loss is combined with the feature reconstruction loss to update our regressor (Equation~\ref{eq:loss_g}).  The adversarial loss of the discriminator is defined as
\begin{equation}
\label{eq:loss_adv_d}
\mathcal{L}_{adv\!\_D}=-\frac{1}{b}\sum_{i=0}^{b-1}(\log{D_{\upsilon}}(I_i) + \log(1- D_{\upsilon}(\hat{I}_i))),
\end{equation}
which estimates the divergence between the distribution of the generated images and the ground truth images.

\subsection{Techniques to Stabilize Training}
\label{sec:stabilize}

We use several techniques to stabilize the adversarial training of $R_{\omega}$ and $D_{\upsilon}$.  The instability of adversarial trainings is a well-known problem~\cite{GoodfellowAMXFOCB14}, especially when the resolution of synthesized images is high~\cite{BergerLL19}.
The previous work~\cite{BergerLL19} divided the training into two stages for stabilization.  In the first stage, the opacity GAN that produces $64{\times}64$ opacity images is trained, whereas the opacity-to-color translation GAN is trained in the second stage to produce $256{\times}256$ color images, conditioning on the $64{\times}64$ opacity images.
In this work, we train a single pair of adversarial networks (i.e., $R_{\omega}$ and $D_{\upsilon}$) that directly produces $256{\times}256$ color images with the help of recent techniques in stabilizing the adversarial training, including the spectral normalization~\cite{MiyatoKKY18} and the two time-scale update rule (TTUR)~\cite{MartinRUNH17}.

\subsubsection{Spectral Normalization}

Spectral normalization~\cite{MiyatoKKY18} is used to mitigate the instability of the discriminator, which is a major challenge in stabilizing the adversarial training.  Spectral normalization is a weight normalization technique, which outperforms other weight normalization techniques in many image synthesis tasks as shown in~\cite{KurachLZMG18}.  Spectral normalization normalizes the weight matrix of each layer based on the first singular value of the matrix.  With spectral normalization, the discriminator is enforced to be Lipschitz continuous, such that the discriminator is constrained and stabilized to some extent.  Spectral normalization is applied on each layer of the discriminator without changing the network architecture; hence spectral normalization is not labeled in Figure~\ref{fig:discriminator}.

\subsubsection{Learning Rate}

The learning rates of $R_{\omega}$ and $D_{\upsilon}$ are critical for the stability of the adversarial training.  This work uses the Adam optimizer~\cite{KingmaB15} that changes the learning rate of each weight dynamically during training with respect to the momentum of the weight gradients.  In detail, the learning rate in the Adam optimizer is controlled by three hyperparmeters: the initial learning rate $\alpha$, the first-order momentum $\beta_1$, and the second-order momentum $\beta_2$.  To stabilize the training, a small $\alpha$ is often preferred; and we found that $5\times10^{-5}$ stabilized the training in our cases.  In addition, we found that a bigger $\beta_1$ often cripples the training and set $\beta_1$ to $0$ as suggested in~\cite{ZhangGMO18, BrockDS18}.  Compared with $\beta_1$, $\beta_2$ has less influence on the stability of the training, which is set to $0.999$.

In previous works on training GANs, we found that people often update the discriminator more frequently than the generator, because they do not want to update the generator based on a discriminator that is not strong enough.  Doing so, however, leads to a longer training time.  Our work uses the same update frequency for the regressor and discriminator but with different learning rates $\alpha_D$ and $\alpha_R$ (i.e., the TTUR technique~\cite{MartinRUNH17}).  Based on the empirical results shown in~\cite{ZhangGMO18, BrockDS18}, we set the learning rate of the discriminator to be 4 times that of the regressor, that is, $\alpha_D=2\times10^{-4}$ and $\alpha_R=5\times10^{-5}$.

\subsection{Training Process}
\vspace{-8pt}

\begin{algorithm}
\caption{Training process of InSituNet.}
\label{alg:training}

\begin{algorithmic}[1]
\Require Training data includes parameters $\left\{P_{sim}, P_{vis}, P_{view}\right\}_{0:N-1}$ and the corresponding images $I_{0:N-1}$.  Initial weights $\omega$ and $\upsilon$ of $R_{\omega}$ and $D_{\upsilon}$, respectively.  The feature comparator $F$.
\Ensure Optimized weights $\omega$ and $\upsilon$
\State Repeat:
\State \quad $\left\{P_{sim}, P_{vis}, P_{view}\right\}_{0:b-1}, I_{0:b-1}$ sampled from training data
\State \quad $\hat{I}_{0:b-1} \gets R_{\omega}(\left\{P_{sim}, P_{vis}, P_{view}\right\}_{0:b-1})$
\State \quad $\upsilon \gets \text{Adam}( \,\nabla_{\upsilon} \mathcal{L}_{adv\!\_D}( \,I_{0:b-1}, \hat{I}_{0:b-1}; \upsilon) \,, \upsilon, \alpha_D, \beta_{1}, \beta_{2}) \,$
\State \quad $\omega \gets \text{Adam}( \,\nabla_{\omega} \mathcal{L}( \,I_{0:b-1}, \hat{I}_{0:b-1}; \omega) \,, \omega, \alpha_R, \beta_{1}, \beta_{2}) \,$
\State Until exit criterion is satisfied
\end{algorithmic}
\end{algorithm}
\vspace{-8pt}

The process of training our regressor and discriminator is shown in Algorithm~\ref{alg:training}.  Given the training data collected in situ, namely, $N$ pairs of paramters $\left\{P_{sim}, P_{vis}, P_{view}\right\}_{0:N-1}$ and the corresponding images $I_{0:N-1}$, we first initialize the network weights $\omega$ and $\upsilon$ using the orthogonal initialization~\cite{SaxeMG13}.
Then, the discriminator and regressor are updated alternatively by using the stochastic gradient descent until the exit criterion is satisfied.  The exit criterion used in this work is the maximum number of iterations, which is set to 125,000 because the loss converged in our cases after 125,000 iterations.

In each iteration, a batch of parameters $\left\{P_{sim}, P_{vis}, P_{view}\right\}_{0:b-1}$ and the corresponding images $I_{0:b-1}$ are sampled from the training data (line 2), where $b$ is the batch size.  Next, the current $R_{\omega}$ takes $\left\{P_{sim}, P_{vis}, P_{view}\right\}_{0:b-1}$ as inputs and produces $\hat{I}_{0:b-1}$ (line 3).  According to the loss $\mathcal{L}_{adv\!\_D}$ defined on $I_{0:b-1}$ and $\hat{I}_{0:b-1}$ in Equation~\ref{eq:loss_adv_d}, the weights of the discriminator are updated (line 4).  Similarly, the weights of the regressor are updated as well, according to the loss function $\mathcal{L}$ (defined in Equations~\ref{eq:loss_g},~\ref{eq:loss_perc}, and~\ref{eq:loss_adv_g}), which is computed using the feature comparator $F$ and the updated discriminator $D_{\upsilon}$ (line 5).  When updating the weights $\upsilon$ and $\omega$, the gradients $\nabla_{\upsilon}$ and $\nabla_{\omega}$ of the loss functions $\mathcal{L}_{adv\!\_D}$ and $\mathcal{L}$ are computed, respectively.  With $\nabla_{\upsilon}$ and $\nabla_{\omega}$, the weights $\upsilon$ and $\omega$ are updated through two Adam optimizers using the learning rates discussed in the preceding section.

\vspace{-4pt}
\section{Parameter Space Exploration with InSituNet}
\label{sec:vis_sys}
\vspace{-10pt}

\setlength{\abovecaptionskip}{4pt}
\setlength{\belowcaptionskip}{-10pt}
\begin{figure}[tbh]
  \centering
  \includegraphics[width=\linewidth]{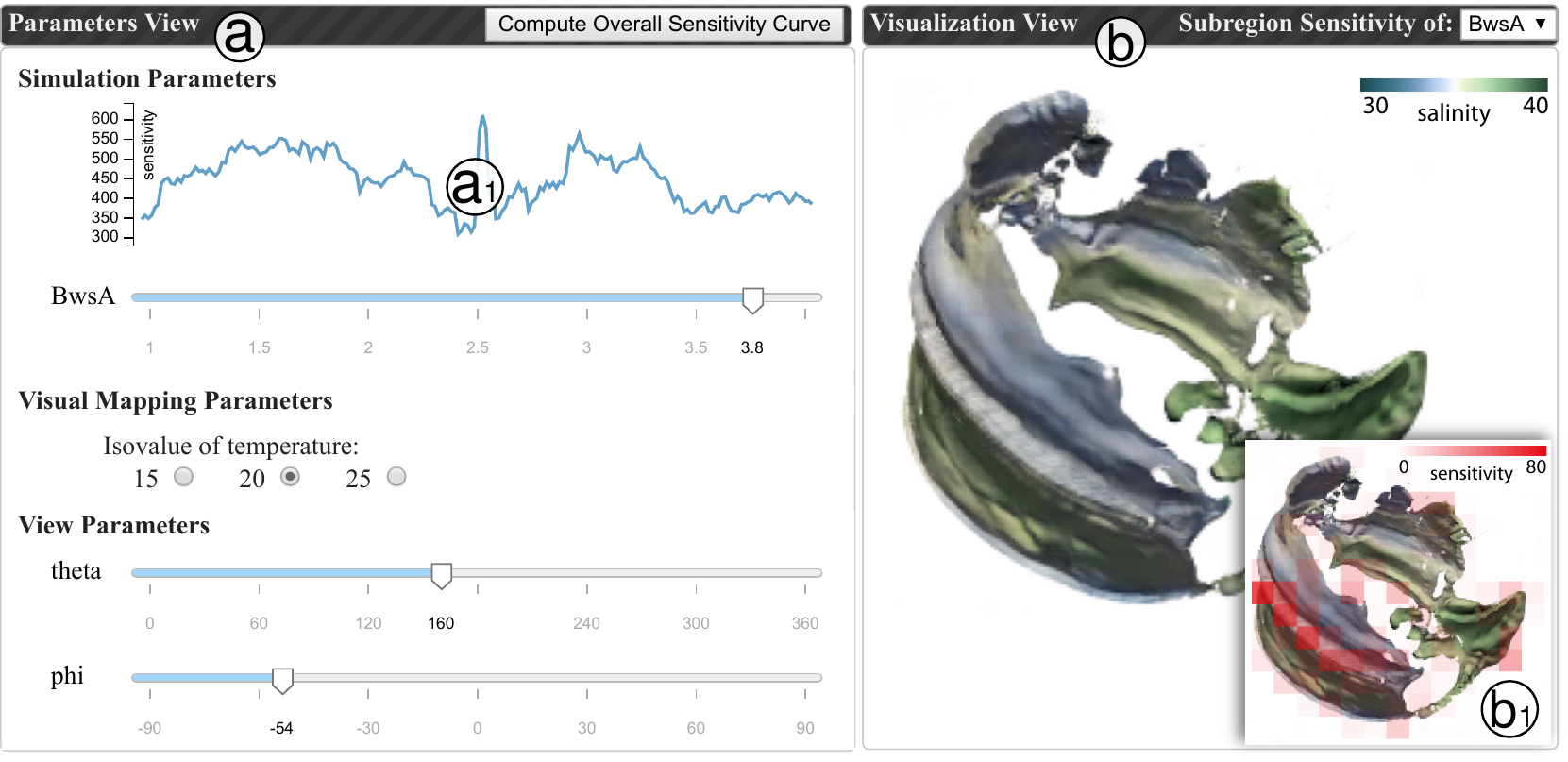}
  \caption{Visual interface for parameter space exploration.  (a) The three groups of parameters: simulation, visual mapping, and view parameters.  (b) The predicted visualization image and the sensitivity analysis result.
  }
  \label{fig:gui}
\end{figure}
\setlength{\abovecaptionskip}{10pt}
\setlength{\belowcaptionskip}{0pt}

With the trained InSituNet, users can perform parameter space exploration of ensemble simulations from two perspectives.  First, with InSituNet's forward propagations, users can interactively infer the visualization results for arbitrary parameters within the parameter space.  Second, using InSituNet's backward propagations, users can investigate the sensitivity of different parameters and thus have better understanding on parameter selections.  To support the parameter space exploration, we built an interactive visual interface as shown in Figure~\ref{fig:gui}, which contains two views: Parameters View (Figure~\ref{fig:gui}(a)) and Visualization View (Figure~\ref{fig:gui}(b)).  In the following, we explain how users can perform parameter space exploration with this visual interface.

\begin{table*}[tbh]
  \caption{Datasets and timings: $k$ controls the size of InSituNet to cope with datasets in different complexities; diversity~\cite{8320546} measures how diverse the generated images are; $t_{sim}$, $t_{vis}$, and $t_{tr}$ are timings for running ensemble simulations, visualizing data in situ, and training InSituNet, respectively; $t_{fp}$ and $t_{bp}$ are timings for a forward and backward propagation of the trained InSituNet, respectively.}
  \label{tab:datasets}
  \centering
  \setlength{\tabcolsep}{2pt}
  \vspace*{-1mm}
  {\small
  \begin{tabular}{l|cc|c|cc|c|c|ccc|ccccc}
  \multirow{2}{*}{Simulation} & \multicolumn{2}{c|}{$P_{sim}$} & \multirow{2}{*}{$P_{vis}$} & \multicolumn{2}{c|}{$P_{view}$} & \multirow{2}{*}{$k$} & \multirow{2}{*}{Diversity} & \multicolumn{3}{c|}{Size (GB)} & \multicolumn{5}{c}{Performance} \\
  & Name & Number & & Name & Number & & & Raw & Image & Network & $t_{sim}$ (hr) & $t_{vis}$ (hr) & $t_{tr}$ (hr) & $t_{fp}$ (s) & $t_{bp}$ (s) \\
  \hline
  SmallPoolFire & $Ck, C$ & 4,000 & pseudo-coloring with 5 color schemes & N/A & N/A & 32 & 2.72 & $\approx$25.0 & 0.43 & 0.06 & 1,420.0 & 6.45 & 16.40 & 0.031 & 0.19 \\
  Nyx & $OmM, OmB, h$ & 500 & volume rendering with a transfer function & $\theta,\phi$ & 100 & 48 & 1.72 & $\approx$30.0 & 3.92 & 0.12 & 537.5 & 8.47 & 18.02 & 0.033 & 0.22 \\
  MPAS-Ocean & $BwsA$ & 300 & isosurface visualization with 3 isovalues & $\theta,\phi$ & 100 & 48 & 1.75 & $\approx$300.0 & 3.46 & 0.15 & 229.5 & 10.73 & 18.13 & 0.033 & 0.23 \\
  \end{tabular}
  }
\vspace*{-6mm}
\end{table*}

\subsection{Inference of Visualization Results}

InSituNet is able to interactively infer the visualization results for any user-selected parameter values.  As shown in Figure~\ref{fig:gui}(a), the three groups of input parameters for the InSituNet are visualized by using different GUI widgets.  For the simulation and view parameters, because their values are usually in continuous ranges, we visualize them using slider bars whose ranges are clipped to the corresponding parameters' predefined value ranges.  Users are able to select arbitrary parameter values by interacting with those sliders.  For the visual mapping parameters, users can switch among a set of predetermined options using the ratio buttons, for example, selecting different isovalues for isosurface visualizations, as shown in Figure~\ref{fig:gui}(a).

The selected values for the three groups of parameters are fed into the trained InSituNet.  Through a forward propagation of the network, which takes around 30 ms, the corresponding visualization image for the given set of parameters is generated and visualized in the Visualization View, as shown in Figure~\ref{fig:gui}(b).

\subsection{Sensitivity Analysis on Simulation Parameters}

Because InSituNet is differentiable, users can perform sensitivity analysis for the simulation parameters using the network's backward propagations.  Specifically, users can compute the derivative of a scalar value derived from the generated image (e.g., $L_1$ norm of the pixel values) with respect to a selected simulation parameter.  The absolute value of the derivative can be treated as the sensitivity of the parameter, which indicates how much the generated image will change if the parameter gets changed.  Note that the sensitivity analysis in this work is used to reflect the changes (with respect to the parameters) in the image space rather than the data space.  Inspired by~\cite{BergerLL19}, our analysis includes overall sensitivity analysis and subregion sensitivity analysis.

In overall sensitivity analysis, we focus on analyzing the sensitivity of the entire image with respect to each simulation parameter across its value range.  To this end, we sweep each parameter across its value range while fixing the values of other parameters.  Images are then generated from the selected parameter values and aggregated into a scalar (i.e., the $L_1$ norm of the pixel values).  The aggregated scalar values are then back propagated through the InSituNet to obtain the sensitivity of the selected parameter values.  In the end, a list of sensitivity values is returned for each parameter and visualized as a line chart on top of the slider bar corresponding to the parameter (Figure~\ref{fig:gui}(a1)) to indicate how sensitive the parameter is across its value range.

In subregion sensitivity analysis, we analyze the sensitivity of a selected parameter for different subregions of the generated image.  This analysis is done by partitioning the visualization image into blocks and computing the sensitive of the parameter for the $L_1$ norm of the pixel values in each block.  The computed sensitivity values are then color coded from white to red and overlaid on top of the visualization image to indicate what regions are more sensitive with respect to the selected parameter (red blocks in Figure~\ref{fig:gui}(b1)).

\section{Results}
\label{sec:results}

We evaluated InSituNet using combustion, cosmology, and ocean simulations (Section~\ref{sec:datasets}) from four aspects: (1) providing implementation details and analyzing performance (Section~\ref{sec:performance}); (2) evaluating the influence of different hyperparameters (Section~\ref{sec:validation}); (3) comparing with alternative methods (Section~\ref{sec:comparison}); and (4) performing parameter space exploration and analysis with case studies (Section~\ref{sec:parameter_analysis}).

\subsection{Ensemble Simulations}
\label{sec:datasets}

We evaluated the proposed approach using three ensemble simulations: SmallPoolFire~\cite{WellerTJF98}, Nyx~\cite{AlmgrenBLLA13}, and MPAS-Ocean~\cite{RinglerPHJJM13}.  They are summarized in Table~\ref{tab:datasets} (left) and detailed below.

\textbf{SmallPoolFire} is a 2D combustion simulation from the OpenFOAM simulation package~\cite{WellerTJF98}.  We used it as a test case to evaluate InSituNet by studying two parameters: a turbulence parameter $Ck{\in}[0.0925, 0.0975]$ and a combustion parameter $C{\in}[4.99, 5.01]$.  We sampled 4,000 parameter settings from the parameter space: 3,900 for training and 100 for testing.  Images were generated for the temperature field by using pseudo-coloring with five predefined color schemes.  To study how diverse the generated images are, we use the method proposed by Wang et al.~\cite{8320546}, which measures the diversity as the reciprocal of the average structural similarity (SSIM) between every pair of images.  The diversity of the images in this dataset is 2.72, which means the average SSIM is smaller than 0.4.

\textbf{Nyx} is a cosmological simulation developed by Lawrence Berkeley National Laboratory.  Based on the scientists' suggestion, we studied three parameters: the total matter density ($OmM\in[0.12, 0.155]$), the total density of baryons ($OmB\in[0.0215, 0.0235]$), and the Hubble constant ($h\in[0.55, 0.85]$).  We sampled 500 parameter settings from the parameter space: 400 for training and 100 for testing.  The simulation was conducted with each parameter setting and generated a $256{\times}256{\times}256$ volume representing the log density of the dark matters.  The volume was visualized in situ by using volume rendering with a predefined transfer function of the wave colormap\footnote{https://sciviscolor.org} and from 100 different viewpoints.  The diversity of the generated images is 1.72.

\textbf{MPAS-Ocean} is a global ocean simulation developed by Los Alamos National Laboratory.  Based on the domain scientists' interest, we studied the parameter that controls the bulk wind stress amplification ($BwsA\in[1, 4]$).  We generated 300 ensemble members with different $BwsA$ values. We used 270 of them for training and the rest for testing.  The isosufaces of the temperature field (with isovalue=\{15, 20, 25\}) were extracted and visualized from 100 different viewpoints for each ensemble member.  The isosurfaces were colored based on salinity, using the colormap suggested by Samsel et al.~\cite{SamselPATRA15}.  The diversity of the generated images is 1.75.

\subsection{Implementation and Performance}
\label{sec:performance}

The proposed approach consists of three components: the in situ data collection, the training of InSituNet, and the visual exploration and analysis component.  We discuss the implementation details and performance of the three components in the following.

The in situ visualization was implemented by using ParaView Catalyst\footnote{https://www.paraview.org/in-situ} following the Cinema framework~\cite{AhrensJOPRP14, AhrensJOPRFBPSB14}.  The simulations and in situ visualization were conducted on a supercomputer of 648 computation nodes.  Each node contains an Intel Xeon E5-2680 CPU with 14 cores and 128 GB of main memory.  We used 1, 28, and 128 processes, respectively, for the SmallPoolFire, Nyx, and MPAS-Ocean simulations.  InSituNet was implemented in PyTorch\footnote{https://pytorch.org} and trained with an NVIDIA DGX-1 system, which contains 8 NVIDIA V100 GPUs with NVlink.  The visual interface was implemented based on a web server/client framework.  The interface was implemented with D3.js on the client side, and the images were generated from a Python server (with the assist of the trained InSituNet) and sent to the client for visualization.  The visual exploration and analysis were tested on a desktop with an Intel Core i7-4770 CPU and an NVIDIA 980Ti GPU.

The space and computation costs using the proposed approach for the three different datasets are listed in Table~\ref{tab:datasets} (right).  The size of InSituNet is less than 1\% and 15\% of the raw simulation data and the image data, respectively.  In terms of data reduction, we also compare our approach with several data compression methods and the results can be found in the supplementary material.  The training of InSituNet generally takes more than 10 hours, but the time is much less than actually running the ensemble simulations with extra parameter settings.  After training, a forward or backward propagation of InSituNet takes less than one second on a single NVIDIA 980Ti GPU.

\subsection{Model Evaluation for Different Hyperparameters}
\label{sec:validation}

We evaluated InSituNet trained with different hyperparameters (i.e., loss functions, network architectures, and numbers of training samples) qualitatively and quantitatively using the data that were excluded from the training to study two questions:  (1) Is InSituNet able to generate images that are close to the ground truth images?  (2) How do the choices of hyperparameters influence the training results?

For quantitative evaluations, we used four metrics that focus on different aspects to compare the predicted images with the ground truth images, including peak signal-to-noise ratio (PSNR), SSIM, earth mover's distance (EMD) between color histograms~\cite{BergerLL19}, and Fr\'{e}chet inception distance (FID)~\cite{MartinRUNH17}.

\textbf{PSNR} measures the pixel-level difference between two images using the aggregated mean squared error between image pixels.  A higher PSNR indicates that the compared images are more similar pixel wise.

\textbf{SSIM} compares two images based on the regional aggregated statistical information (e.g., mean and standard deviation of small patches) between them.  A higher SSIM means the compared images are more similar from a structural point of view.

\textbf{EMD} is used in~\cite{BergerLL19} to quantify the distance between the color histograms of two images.  A lower EMD means the compared images are more similar according to their color distributions.

\textbf{FID} approximates the distance between two distributions of images, which is widely used in recent image synthesis works~\cite{ZhangGMO18, BrockDS18} as a complementary to other metrics.  A lower FID suggests the two image collections are more similar statistically.

\subsubsection{Loss Functions}

\setlength{\abovecaptionskip}{4pt}
\setlength{\belowcaptionskip}{-19pt}
\begin{figure}[tb]
  \centering
  \includegraphics[width=\linewidth]{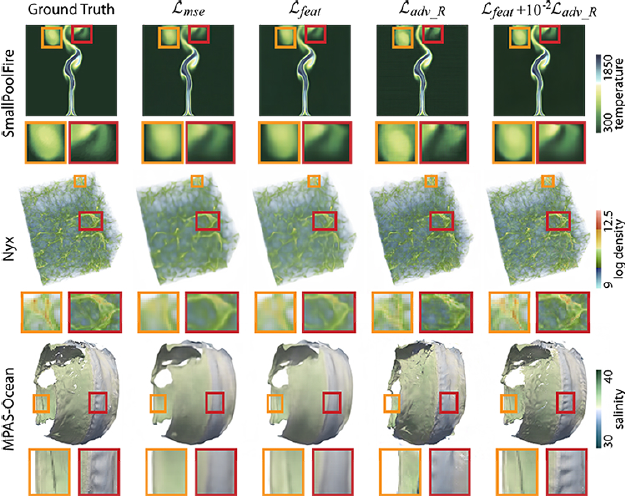}
  \caption{Qualitative comparison of InSituNet trained with different loss functions.  Combining $\mathcal{L}_{feat}$ and $\mathcal{L}_{adv\!\_R}$ gives the results of high quality.
  }
  \label{fig:eval_losses}
\end{figure}
\setlength{\abovecaptionskip}{10pt}
\setlength{\belowcaptionskip}{0pt}

We evaluated InSituNet trained with different loss functions including the mean squared error $\mathcal{L}_{mse}$, the feature reconstruction loss $\mathcal{L}_{feat}$, the adversarial loss $\mathcal{L}_{adv\!\_R}$, and the combination of $\mathcal{L}_{feat}$ and $\mathcal{L}_{adv\!\_R}$.

Figure~\ref{fig:eval_losses} compares the images generated by InSituNet trained with different loss functions with the ground truth (The enlarged figure can be found in the supplementary material).  We can see that using $\mathcal{L}_{mse}$ often generates over-smoothed images lacking high-frequency features, whereas using $\mathcal{L}_{feat}$ can mitigate the problem to some extent.  Using $\mathcal{L}_{adv\!\_R}$ can generate images that are as sharp as the ground truth, but the features are often not introduced in the desired positions.  By combining $\mathcal{L}_{feat}$ and $\mathcal{L}_{adv\!\_R}$, we are able to generate images with sharp features, and those images are also similar to the ground truth.

\begin{table}[tb]
  \caption{Quantitative evaluation of InSituNet trained with different loss functions.  The model trained with the combination of $\mathcal{L}_{feat}$ and $\mathcal{L}_{adv\!\_R}$ generates images with the best EMD and FID and only a slightly lower PSNR and SSIM compared with the model trained with $\mathcal{L}_{mse}$ or $\mathcal{L}_{feat}$.}
  \label{tab:eval_losses}
  \centering
  {\small
  \begin{tabular}{lc|c|c|c|c}
  \multicolumn{2}{c|}{} & $\mathcal{L}_{mse}$ & $\mathcal{L}_{feat}$ & $\mathcal{L}_{adv\!\_R}$ & $\mathcal{L}_{feat} + 10^{-2}\mathcal{L}_{adv\!\_R}$ \\
  \hline
  \multirow{4}{*}{SmallPoolFire} & PSNR & \textbf{25.090} & 24.937 & 20.184 & 24.288 \\
                                 & SSIM & 0.9333 & \textbf{0.9390} & 0.8163 & 0.9006 \\
                                 & EMD  & 0.0051 & 0.0064 & 0.0056 & \textbf{0.0037} \\
                                 & FID  & 21.063 & 15.881 & 12.859 & \textbf{9.4747} \\
  \hline
  \multirow{4}{*}{Nyx} & PSNR & \textbf{31.893} & 29.055 & 24.592 & 29.366 \\
                       & SSIM & 0.8684 & \textbf{0.8698} & 0.7081 & 0.8336 \\
                       & EMD  & 0.0037 & 0.0083 & 0.0064 & \textbf{0.0022} \\
                       & FID  & 60.825 & 54.670 & 24.036 & \textbf{6.2694} \\
  \hline
  \multirow{4}{*}{MPAS-Ocean} & PSNR & \textbf{26.944} & 26.267 & 17.099 & 24.791 \\
                              & SSIM & \textbf{0.8908} & 0.8885 & 0.7055 & 0.8655 \\
                              & EMD  & 0.0025 & 0.0044 & 0.0036 & \textbf{0.0017} \\
                              & FID  & 115.74 & 120.37 & 28.927 & \textbf{21.395} \\
  \end{tabular}
  }
\vspace*{-6mm}
\end{table}

Table~\ref{tab:eval_losses} reports the quantitative results from using different loss functions.  We found that using $\mathcal{L}_{mse}$ gives the best PSNR, because the network using $\mathcal{L}_{mse}$ is trained to minimize the mean squared error (i.e., maximize the PSNR).  Using $\mathcal{L}_{feat}$ gives the best SSIM in some cases, because it focuses more on the structure of the images.  However, using $\mathcal{L}_{mse}$ or $\mathcal{L}_{feat}$ often results in poor performance regarding EMD and FID.  When training InSituNet with $\mathcal{L}_{adv\!\_R}$, the FID value can be improved, but the values of PSNR and SSIM drop a lot.  By combining $\mathcal{L}_{feat}$ and $\mathcal{L}_{adv\!\_R}$, both EMD and FID improved a lot, though the PSNR and SSIM got slightly worse than using $\mathcal{L}_{mse}$ or $\mathcal{L}_{feat}$.

\vspace{-12pt}
\setlength{\abovecaptionskip}{2pt}
\setlength{\belowcaptionskip}{-6pt}
\begin{figure}[tbh]
  \centering
  \includegraphics[width=.8\linewidth]{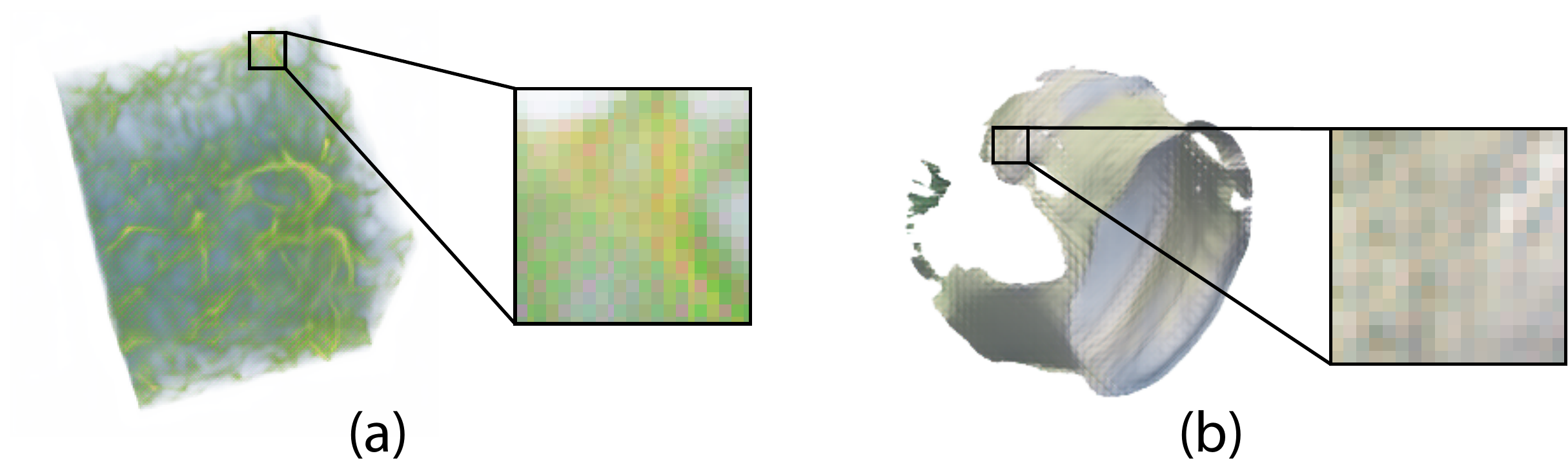}
  \caption{Images generated by InSituNet trained with $\mathcal{L}_{feat}$ that uses different layers after the first pooling layer of VGG-19: (a) \texttt{relu2\_1} and (b) \texttt{relu3\_1}.  Checkerboard artifacts are introduced.
  }
  \label{fig:eval_perc_layers}
\end{figure}
\setlength{\abovecaptionskip}{10pt}
\setlength{\belowcaptionskip}{0pt}

For $\mathcal{L}_{feat}$, using which layer of the pretrained VGG-19 to extract features from images can affect the image synthesis results.  Through empirical studies, we found that using any layers after the first pooling layer of VGG-19 will introduce undesired checkerboard artifacts, because of the ``inhomogeneous gradient update'' of the pooling layer~\cite{AitkenLTCWS17}, as shown in Figure~\ref{fig:eval_perc_layers}.  Hence, we use the last layer right before the first pooling layer, which is the layer \texttt{relu1\_2}.

\vspace{-6pt}
\begin{table}[tbh]
  \caption{Evaluating the weight $\lambda$ of $\mathcal{L}_{adv\!\_R}$: $\lambda=0.01$ provides the results that balance the PSNR, SSIM, EMD, and FID.}
  \label{tab:eval_adv_weight}
  \centering
  \setlength{\tabcolsep}{10pt}
  {\small
  \begin{tabular}{l|c|c|c|c}
  & $\lambda=0.005$ & $\lambda=0.01$ & $\lambda=0.02$ & $\lambda=0.04$ \\
  \hline
  PSNR & \textbf{30.043} & 29.366 & 29.040 & 27.232 \\
  SSIM & \textbf{0.8619} & 0.8336 & 0.8253 & 0.7680 \\
  EMD & 0.0041 & \textbf{0.0022} & 0.0023 & 0.0025 \\
  FID & 21.267 & \textbf{6.2694} & 6.6819 & 9.8992 \\
  \end{tabular}
  }
\vspace*{-3mm}
\end{table}

We also evaluated the influence of the weight $\lambda$ for $\mathcal{L}_{adv\!\_R}$ when combining it with $\mathcal{L}_{feat}$ (defined in Equation~\ref{eq:loss_g}), and the results are shown in Table~\ref{tab:eval_adv_weight}.  We found that increasing $\lambda$ over 0.01 cannot improve the accuracy of the generated images any further.  In addition, a small $\lambda$ (i.e., 0.005) will hurt the image accuracy in terms of EMD and FID, although the value of PSNR and SSIM can be improved slightly.  We thereby set $\lambda$ to 0.01 to balance its effects on the four metrics.

\subsubsection{Network Architectures}

\setlength{\abovecaptionskip}{2pt}
\setlength{\belowcaptionskip}{-10pt}
\begin{figure}[tbh]
  \centering
  \includegraphics[width=.8\linewidth]{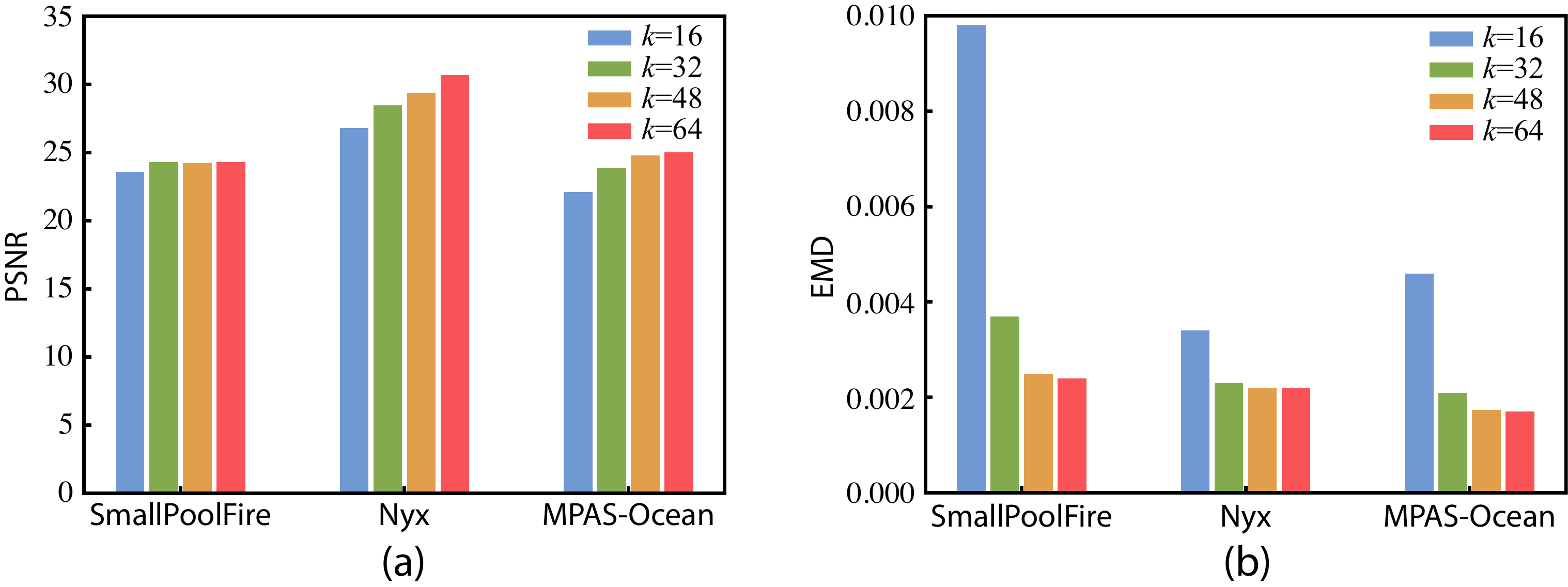}
  \caption{Quantitative evaluation of different network architectures controlled by $k$ with (a) PSNR and (b) EMD.
  }
  \label{fig:eval_k}
\end{figure}
\setlength{\abovecaptionskip}{10pt}
\setlength{\belowcaptionskip}{0pt}

\begin{table}[tbh]
  \caption{Size and training time of different network architectures controlled by $k$ for the Nyx dataset.}
  \label{tab:eval_k}
  \centering
  \setlength{\tabcolsep}{10pt}
  \vspace*{-1mm}
  {\small
  \begin{tabular}{l|c|c|c|c}
  & $k=16$ & $k=32$ & $k=48$ & $k=64$ \\
  \hline
  Network Size (MB) & 26.4 & 67.4 & 125.2 & 199.6 \\
  Training Time (hr) & 13.73 & 16.42 & 18.02 & 20.17 \\
  \end{tabular}
  }
\vspace*{-6mm}
\end{table}

We evaluated InSituNet with different network architectures in terms of the accuracy of predicted images, the network size, and the training time.  As mentioned in Section~\ref{sec:architecture}, the architecture of our network is controlled by a constant $k$, which controls the number of convolutional kernels in the intermediate layers.  In this experiment, we evaluated four $k$ values: 16, 32, 48, and 64.

Figure~\ref{fig:eval_k} shows the PSNR and EMD of images generated by InSituNet with the four $k$ values.  We can see that InSituNet with larger $k$ values can generate more (or at least equally) accurate images, because a larger $k$ gives more expressive power to the neural network.  On the other hand, training InSituNet with a larger $k$ also costs more time, and more storage will be needed to store the networks, as shown in Table~\ref{tab:eval_k} using the Nyx dataset as an example.  Hence, to balance the accuracy of the generated images and the cost from both computation and storage, we set $k$ to 32, 48, and 48 for the SmallPoolFire, Nyx, and MPAS-Ocean, respectively.

\subsubsection{Number of Ensemble Runs used for Training}

\vspace{-6pt}
\begin{table}[tbh]
  \caption{Evaluation of the number of ensemble runs used for training.}
  \label{tab:eval_sn}
  \centering
  {\small
  \begin{tabular}{l|c|c|c|c|c}
  Simulation & \# Ensemble Runs & PSNR & SSIM & EMD & FID \\
  \hline
  \multirow{4}{*}{SmallPoolFire} & 900  & 21.842 & 0.8714 & 0.0040 & 14.398 \\
                                 & 1900 & 23.192 & 0.9016 & 0.0036 & 11.732 \\
                                 & 2900 & 23.932 & 0.9018 & 0.0037 & 9.5813 \\
                                 & 3900 & 24.288 & 0.9006 & 0.0037 & 9.4747 \\
  \hline
  \multirow{4}{*}{Nyx} & 100 & 28.108 & 0.7951 & 0.0025 & 9.8818 \\
                       & 200 & 29.404 & 0.8319 & 0.0022 & 6.5481 \\
                       & 300 & 29.398 & 0.8326 & 0.0023 & 6.4239 \\
                       & 400 & 29.366 & 0.8336 & 0.0022 & 6.2694 \\
  \hline
  \multirow{4}{*}{MPAS-Ocean} & 70  & 24.347 & 0.8554 & 0.0016 & 37.229 \\
                              & 140 & 24.593 & 0.8607 & 0.0017 & 28.380 \\
                              & 210 & 24.732 & 0.8643 & 0.0017 & 22.794 \\
                              & 270 & 24.791 & 0.8655 & 0.0017 & 21.395 \\
  \end{tabular}
  }
\vspace*{-2mm}
\end{table}

We compared InSituNet trained using different numbers of ensemble runs (Table~\ref{tab:eval_sn}) to study how many ensemble runs will be needed to train a good model for the three simulations. We found that this number is different in different simulations, depending on the complexity of the mapping between simulation parameters and visualization results.  Experiment results show that the accuracy of generated images becomes stable when the number of ensemble runs is greater than 2,900, 200, and 210 for the SmallPoolFire, Nyx, and MPAS-Ocean simulation, respectively.  As a result, we used 3,900, 400, and 270 runs from the three simulations to train InSituNet for the rest of the study.

\subsection{Comparison with Alternative Methods}
\label{sec:comparison}

\setlength{\abovecaptionskip}{-6pt}
\setlength{\belowcaptionskip}{-10pt}
\begin{figure}[tbh]
  \centering
  \includegraphics[width=\linewidth]{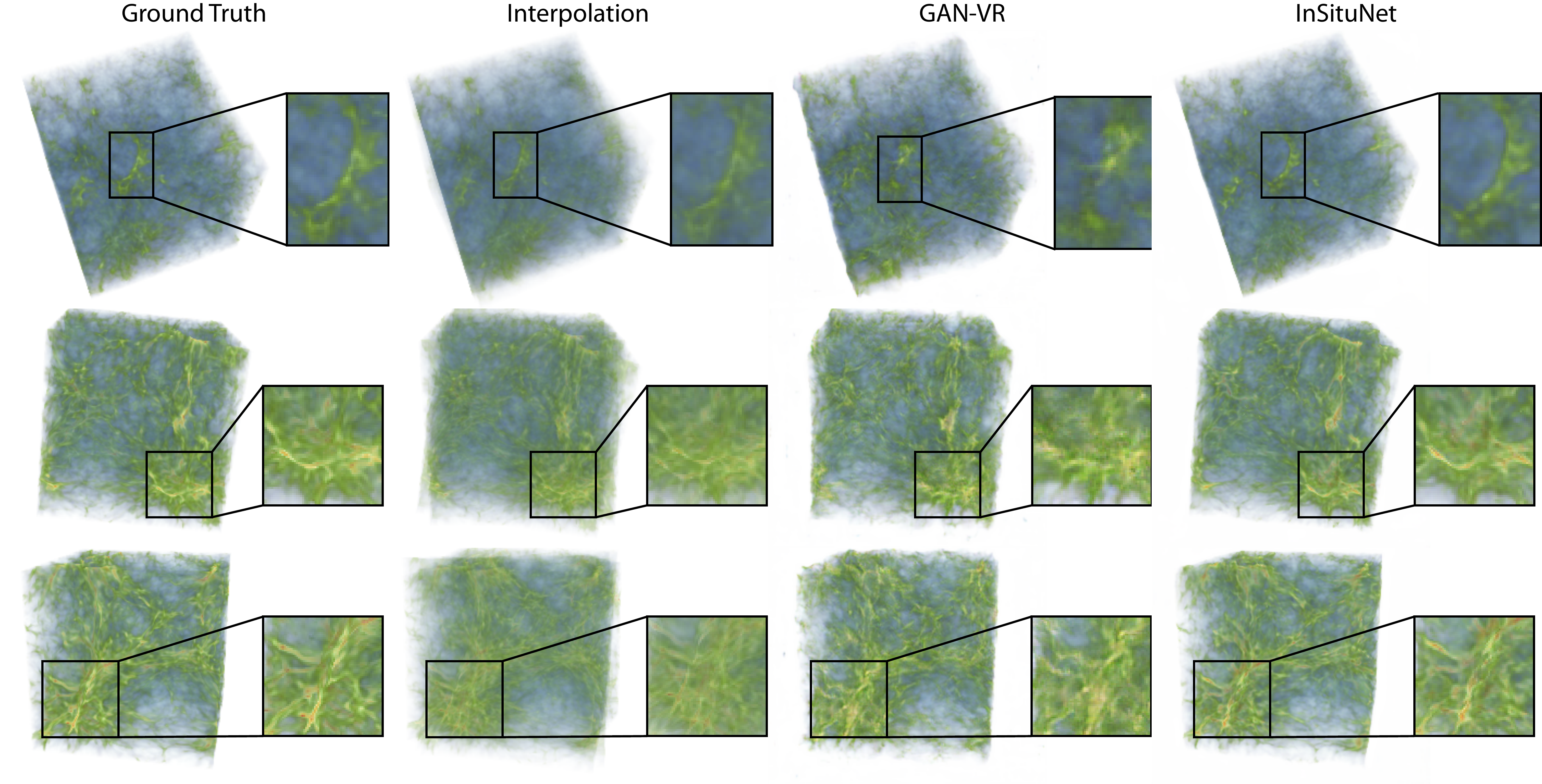}
  \caption{Comparison of the images generated using interpolation, GAN-VR, and InSituNet with the ground truth images.}
  \label{fig:eval_gvr}
\end{figure}
\setlength{\abovecaptionskip}{10pt}
\setlength{\belowcaptionskip}{0pt}

\begin{table}[tbh]
  \caption{Quantitative comparison of images generated with interpolation, GAN-VR, and InSituNet.}
  \label{tab:eval_gvr}
  \centering
  \vspace*{-1mm}
  {\small
  \begin{tabular}{l|c|c|c|c|c}
  & Network Size & PSNR & SSIM & EMD & FID \\
  \hline
  Interpolation & N/A     & 23.932 & 0.6985 & 0.0070 & 58.571 \\
  GAN-VR        & 80.5 MB & 20.670 & 0.6274 & 0.0056 & 38.355 \\
  InSituNet     & \textbf{67.5 MB} & \textbf{28.471} & \textbf{0.8034} & \textbf{0.0023} & \textbf{9.0152} \\
  \end{tabular}
  }
\vspace*{-7mm}
\end{table}

We compared our method with two alternative methods including interpolating images from the training data that close to the target image and the GAN-based volume rendering method (GAN-VR)~\cite{BergerLL19} using the Nyx dataset.  For the interpolation method, we sample $g$ images from the training data whose parameter settings are the top $g$ closest to the parameter setting of the test image and interpolate the sampled images using inverse distance weighting interpolation~\cite{Shepard:1968:TIF:800186.810616}.  We evaluated $g$ from 1 to 5 and present the result of $g=3$ in this section because it balances the four metrics (More results are in the supplementary material).  For GAN-VR, we incorporated the simulation parameters into both the opacity GAN and the opacity-to-color translation GAN and removed the transfer function related parameters because we used a fixed transfer function for this dataset.  For InSituNet, we selected a network architecture whose size is not greater than the size of GAN-VR network, for a fair comparison.

Figure~\ref{fig:eval_gvr} compares the ground truth images with the images generated by using interpolation, GAN-VR, and InSituNet.  With the new network architecture (e.g., the projection-based condition incorporation method in Section~\ref{sec:architecture}), loss functions (e.g., the feature reconstruction loss in Section~\ref{sec:loss}), and training strategies (e.g., the spectral normalization in Section~\ref{sec:stabilize}), InSituNet can generate results that better preserve features compared with the other two methods.  The quantitative comparisons between the three methods are shown in Table~\ref{tab:eval_gvr}.  InSituNet outperforms the other two methods in all four metrics.

\subsection{Parameter Space Exploration}
\label{sec:parameter_analysis}

This section demonstrates the effectiveness of our deep image synthesis driven parameter space exploration through case studies on the Nyx and MPAS-Ocean simulations.

\subsubsection{Case Study with the Nyx Simulation}
\vspace{-8pt}

\setlength{\abovecaptionskip}{4pt}
\setlength{\belowcaptionskip}{-8pt}
\begin{figure}[tbh]
  \centering
  \includegraphics[width=\linewidth]{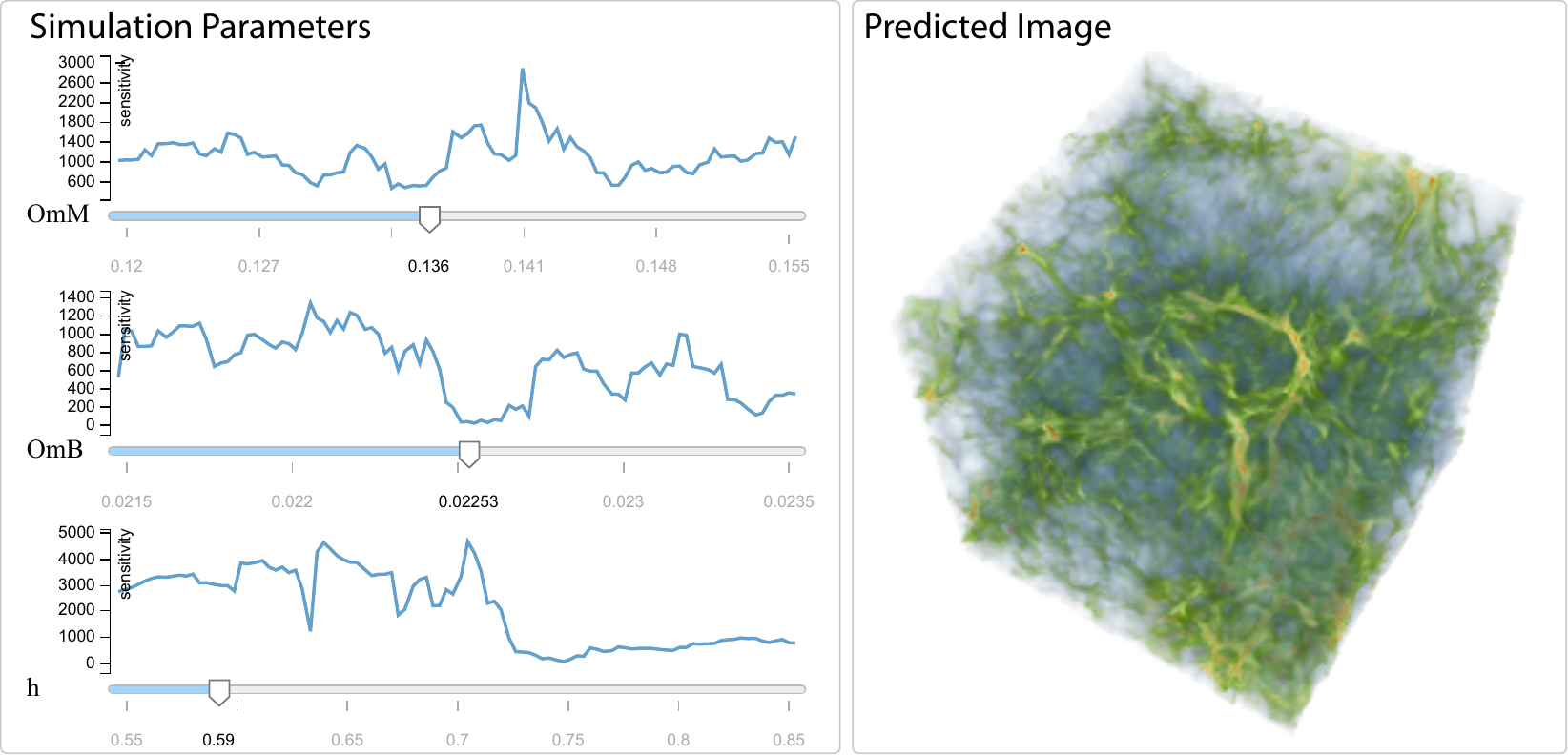}
  \caption{Parameter space exploration with the Nyx simulation.  For the selected parameter values, the sensitivity of different parameters is estimated and visualized as line charts on the left, whereas the predicted image is visualized on the right.}
  \label{fig:case0_0}
\end{figure}
\setlength{\abovecaptionskip}{10pt}
\setlength{\belowcaptionskip}{0pt}

\setlength{\abovecaptionskip}{4pt}
\setlength{\belowcaptionskip}{-18pt}
\begin{figure}[tbh]
  \centering
  \includegraphics[width=\linewidth]{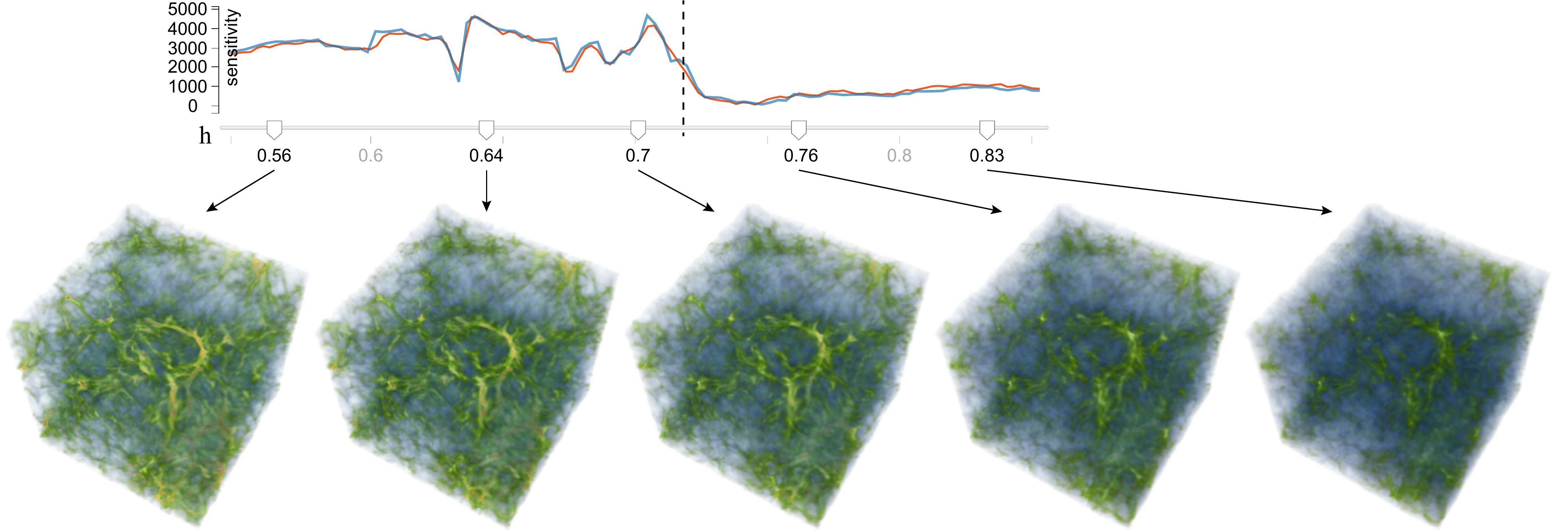}
  \caption{Comparison of the visual appearance of the predicted images using different $h$ values to see the effect of this simulation parameter.}
  \label{fig:case0_1}
\end{figure}
\setlength{\abovecaptionskip}{10pt}
\setlength{\belowcaptionskip}{0pt}

Our first case study is focused on investigating the influence of different simulation parameters (i.e., $OmM$, $OmB$, and $h$) on the Nyx simulation.  The explorations of different visualization settings can be found in our associated video.

Figure~\ref{fig:case0_0} shows a selected parameter setting with the predicted visualization image.  To understand the influence of each parameter, we computed the sensitivity of the three parameters with respect to the $L_1$ norm of the predicted image, shown as the three line charts in Figure~\ref{fig:case0_0}.  From the scale of the three charts (i.e., the values along the vertical axes), we see that parameter $h$ is more sensitive to parameter $OmM$ and parameter $OmM$ is more sensitive to parameter $OmB$.

Focusing on the most sensitive parameter, namely, parameter $h$, we explored how it affects the visual appearance of the predicted images.  Figure~\ref{fig:case0_1} shows five images predicted by using five different $h$ values, while parameter $OmM$ and $OmB$ are fixed at the values shown on the two corresponding slider bars in Figure~\ref{fig:case0_0}.  We first evaluate the accuracy of the sensitivity curve (blue curve in Figure~\ref{fig:case0_1}) computed by backpropagation with the central difference method.  To this end, we first regularly sample the simulation parameters along the curve (128 samples are drawn) and then generate the visualization images with respect to the sampled simulation parameters.  The L1 norm of the generated images is then computed and used to compute the sensitive curve using the central difference method.  The result is shown as the orange curve in Figure~\ref{fig:case0_1}.  We can see that the sensitivity curves generated with the two methods are similar.  From the guidance provided by the line chart in Figure~\ref{fig:case0_1}, we see that parameter $h$ is more sensitive in the first half of its range (i.e., the left side of the dashed line).  The three images generated using $h$ values from this range demonstrate a bigger variance compared with the two images shown on the right (which are images generated by using $h$ values from the second half of its range).

\subsubsection{Case Study with the MPAS-Ocean Simulation}
\vspace{-8pt}

\setlength{\abovecaptionskip}{4pt}
\setlength{\belowcaptionskip}{-18pt}
\begin{figure}[tbh]
  \centering
  \includegraphics[width=.95\linewidth]{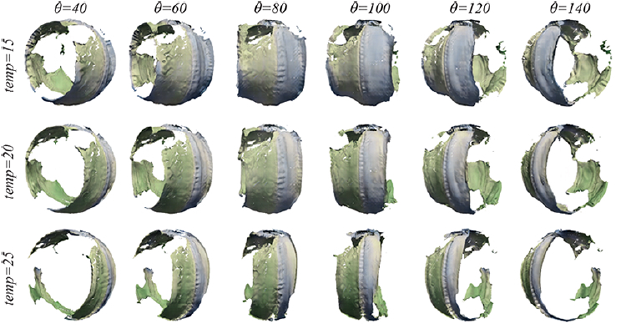}
  \caption{Predicted images of the MPAS-Ocean dataset for different isosurfaces and viewpoints, which reasonably reflect the change of view projections and shading effects.}
  \label{fig:case1_0}
\end{figure}
\setlength{\abovecaptionskip}{10pt}
\setlength{\belowcaptionskip}{0pt}

\setlength{\abovecaptionskip}{2pt}
\setlength{\belowcaptionskip}{-8pt}
\begin{figure}[tbh]
  \centering
  \includegraphics[width=.85\linewidth]{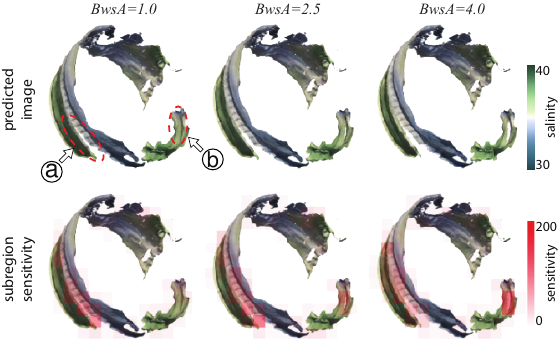}
  \caption{Forward prediction (top row) and backward subregion sensitivity analysis (bottom row) for different $BwsA$.  Regions that influenced by $BwsA$ (i.e., regions a and b) are highlighted by the sensitivity map.}
  \label{fig:case1_1}
\end{figure}
\setlength{\abovecaptionskip}{10pt}
\setlength{\belowcaptionskip}{0pt}

Our next case study explores different parameter settings for the MPAS-Ocean simulation and demonstrates the subregion sensitivity analysis for the simulation parameter $BwsA$, which characterizes the bulk wind stress.  Note that here we focus only on exploration and analysis of new parameter settings, the comparison between the predicted and ground truth images is discussed in the previous sections.

Figure~\ref{fig:case1_0} shows isosurface visualizations of the temperature field with three different isovalues from six different viewpoints.  The value of parameter $BwsA$ is fixed at $1$ in this study.  The images reasonably reflect the change of view projections and shading effects.  More exploration and visualization results can be found in our associated video.

Figure~\ref{fig:case1_1} shows the predicted images when using different $BwsA$ values. All images are generated from the temperature field (isovalue=15) of MPAS-Ocean and from the same viewpoint. The first row of images shows the result of forward inference with different $BwsA$ values, whereas the second row of images overlays the subregion sensitivity maps onto the corresponding images of the first row. The labeled regions (i.e., Figure~\ref{fig:case1_1}a, b) change the most when adjusting the value of $BwsA$, and the subregion sensitivity maps on the second row echo these bigger changes, as indicated by the darker red color.

\section{Limitations, Discussion, and Future Work}
\label{sec:discussion}

This section discusses several directions that we would like to explore in the future: (1) improving the flexibility in exploring arbitrary visual mapping parameters; (2) increasing the accuracy of predicted images; and (3) increasing the resolution of predicted images.

One limitation of our approach is that we restricted the users' ability in exploring arbitrary visual mapping parameters, for example, exhausting all possible transfer functions for volume rendering.  Instead, we allow users to switch only among several predefined visual mappings, for example, the three isovalues when exploring the MPAS-Ocean data.
Theoretically, training a deep learning model to predict visualization images for arbitrary simulation and visualization parameters is feasible.  However, it will require a large number of training images to cover the joint space of all possible simulation and visualization parameters.  For example, in order to train a model that can predict volume rendering results of a single volume data for arbitrary transfer functions, 200,000 training images are required, as shown in~\cite{BergerLL19}.  Consequently, the size of the training data may even exceed the size of the raw simulation data, which offsets the benefit of in situ visualization.
Considering this issue, we would like to explore deep learning techniques that do not require a large number of training samples, such as one- or zero-shot learning, to improve the flexibility of exploration.

Similar to most other machine learning techniques, generating prediction results that are exactly the same as the ground truth is extraordinary difficult.  By taking advantage of recent advances in deep learning for image synthesis, the proposed approach has already outperformed other image synthesis based visualization techniques in terms of the fidelity and accuracy of the generated images (see the comparison in Section~\ref{sec:results}).
However, we believe further improvement is still possible, and we would like explore other network architectures and/or other loss functions to improve our deep image synthesis model.

Our network architecture limits the resolution of output images to $256\times256$, which might not be sufficient for some high-resolution simulation data.  We believe that our network architecture has the potential to generate images with higher resolutions by adding more residual blocks, and we will investigate this approach in the future.

\section{Conclusion}
\label{sec:conclusion}

In this work, we propose InSituNet, a deep learning based image synthesis model supporting the parameter space exploration of large-scale ensemble simulations visualized in situ.  The model is trained to learn the mapping from ensemble simulation parameters to visualizations of the corresponding simulation outputs, conditioned on different visualization settings (i.e., visual mapping and view parameters).
With a trained InSituNet, users can generate visualizations of simulation outputs with different simulation parameters without actually running the expensive simulation, as well as synthesize new visualizations with different visualization settings that are not used during the runs.
Additionally, an interactive visual interface is developed to explore the space of different parameters and investigate their sensitivity using the trained InSituNet.
Through both quantitative and qualitative evaluations, we validated the effectiveness of InSituNet in analyzing ensemble simulations that model different physical phenomena.

\acknowledgments{This work was supported in part by US Department of Energy Los Alamos National Laboratory contract 47145 and UT-Battelle LLC contract 4000159447 program manager Laura Biven.}

\bibliographystyle{abbrv-doi}

\bibliography{template}

\end{document}